\documentclass{emulateapj}

\usepackage{amsmath}
\usepackage{amssymb}
\usepackage{amsfonts}
\usepackage{amstext}
\usepackage{graphicx}
\usepackage{fancybox}
\usepackage[usenames,dvipsnames]{color}
\usepackage{pstricks}
\usepackage{bm}
\usepackage{soul, color}
\usepackage{latexsym}
\usepackage{pifont}
\usepackage{shorttoc}
\usepackage{subfigure}
\usepackage{textcomp} 
\usepackage{float}
\usepackage{enumitem}

\usepackage{bm}
\usepackage{lipsum}
\usepackage{threeparttable}
\usepackage{hyperref}
\hypersetup{urlcolor=cyan}
\def\aap{A\&A}

\def\apjs{ApJS}
\def\apjl{ApJL}

\newcommand{\ha}{H$\alpha$\ }
\newcommand{\lya}{Ly$\alpha$\ }

\newcommand{\angs}{\, {\rm \AA}}

\newcommand{\no}[1]{}
\renewcommand{\exp}[1]{\mathrm{exp}\left(#1\right)}

\newcommand{\myemail}{l.m.ribas@astro.uio.no}
\def\lsim{~\rlap{$<$}{\lower 1.0ex\hbox{$\sim$}}}

\def\gsim{~\rlap{$>$}{\lower 1.0ex\hbox{$\sim$}}}

\shorttitle{SSIM: Extended Ly$\alpha$, \ha and Continuum Emission 
as a Probe of Halo Star Formation}
\shortauthors{Mas-Ribas, Dijkstra, Hennawi, Trenti, Momose \& Ouchi}

\begin{document}

\title{Small-scale Intensity Mapping: Extended L\lowercase{y$\alpha$}, H\lowercase{$\alpha$} and 
Continuum emission\\ as a Probe of Halo Star Formation in High-redshift Galaxies}

\author{Llu\'is Mas-Ribas\altaffilmark{1}} 
\author{Mark Dijkstra\altaffilmark{1}} 
\author{Joseph F. Hennawi\altaffilmark{2,3}}
\author{Michele Trenti\altaffilmark{4}} 
\author{Rieko Momose\altaffilmark{5,6}} 
\author{Masami Ouchi\altaffilmark{5,7}} 

\altaffiltext{1}{Institute of Theoretical Astrophysics, University of Oslo,
Postboks 1029, 0315 Oslo, Norway. \\ 
\url{\myemail}}
\altaffiltext{2}{Max-Planck-Institut f{\"u}r Astronomie, K{\"o}nigstuhl 17, 
D-69117 Heidelberg, Germany} 
\altaffiltext{3}{Department of Physics, University of California, Santa 
Barbara, CA 93106, USA}
\altaffiltext{4}{School of Physics, University of Melbourne, Parkville, VIC 3010, Australia}
\altaffiltext{5}{Institute for Cosmic Ray Research, The University of Tokyo,
5-1-5 Kashiwanoha, Kashiwa, Chiba 277-8582, Japan}
\altaffiltext{6}{Institute of Astronomy, National Tsing Hua University, 101 Section 2 
Kuang-Fu Road, Hsinchu 30013, Taiwan}
\altaffiltext{7}{Kavli Institute for the Physics and Mathematics of the Universe 
(WPI), The University of Tokyo, 5-1-5 Kashiwanoha,
Kashiwa, Chiba 277-8583, Japan}


\begin{abstract}

   Lyman alpha halos are observed ubiquitously around star-forming galaxies at high 
redshift, but their origin is still a matter of debate. We demonstrate that the emission from 
faint unresolved satellite sources, $M_{\rm UV} \gsim -17$, clustered around the central 
galaxies may play a major role in generating spatially extended Ly$\alpha$, continuum 
(${\rm UV + VIS}$) and H$\alpha$ halos. 
We apply the analytic formalism developed in \cite{Masribas2016} to model the halos 
around Lyman Alpha Emitters (LAEs) at $z=3.1$, for several different satellite clustering 
prescriptions. In general, our UV and Ly$\alpha$ surface brightness profiles match the 
observations well at $20\lsim r \lsim 40$ physical kpc from the centers of LAEs. We discuss 
how our profiles depend on various model assumptions and how these can be tested and 
constrained with future H$\alpha$ observations by the James Webb Space Telescope (JWST). 
Our analysis shows how spatially extended halos constrain ({\it i}) the presence of otherwise 
undetectable satellite sources, ({\it ii}) the integrated, volumetric production rates of Ly$\alpha$ and 
LyC photons, and ({\it iii}) their population-averaged escape fractions. These quantities are all directly 
relevant for understanding galaxy formation and evolution and, for high enough redshifts, cosmic 
reionization.

\end{abstract}

\section{Introduction}\label{sec:intro}

   Pioneering studies revealed the presence of diffuse \lya emission 
in the halo of several star-forming galaxies \citep{Moller1998,Fynbo1999,Fynbo2001,
Rauch2008a}. Nowadays, this faint emission is being proved to be nearly ubiquitously in 
galaxies at high redshift, $3\lsim z \lsim5$, by means of stacking analyses  
\citep{Steidel2011,Matsuda2012,Feldmeier2013,Momose2014,Momose2016}, and due to 
the sensitivity and spatial resolution improvement of instruments such as MUSE 
\citep{Bacon2014}. A clear understanding on the origin of these   
extended \lya halos (LAHs; hereafter) is relevant because it yields information 
about the physical conditions of the circumgalactic medium (CGM) and, in turn, on the 
processes governing the formation and evolution of galaxies \citep{Bahcall1969}. 

   The main mechanisms contributing to the existence of LAHs are the cooling of gas 
accreted onto the galaxies, and star formation. Star formation, additionally, can be divided into  
two processes: ({\it i}) The nebular \lya radiation produced in the interstellar medium 
(ISM) diffusing outwards to the CGM via {\it scattering}, and ({\it ii}) the ionizing photons 
escaping the center of the galaxy which produce \lya radiation in the neutral CGM via 
{\it fluorescence}.  

  The \lya cooling radiation produced by the inflowing gas accreted onto the central galaxy has  
been investigated by several authors \citep{Haiman2000, Keres2005,Dekel2006,Shull2009}, 
but the significance of cooling is still difficult to predict accurately and remains uncertain 
\citep{Fardal2001,Yang2006,DijkstraLoeb2009,FaucherGiguere2010,Cantalupo2012,
Rosdahl2012,Lake2015}. The scattering of nebular \lya photons produced in the HII regions of the 
central galaxy likely plays a major role in the observed \lya surface brightness profiles at 
small distances from the center \citep[a few tens of kpc;][ Leclercq et al., in prep.]{Laursen2007,Laursen2009,Steidel2011,Zheng2011,
Wisotzki2015,Xue2017} but, at large impact parameter, scattering from the central galaxy alone usually 
cannot account for the totality of the observed emission \citep[e.g.,][see also 
\citealt{DijkstraKramer2012}]{Lake2015}. Similarly, we demonstrated in \cite{Masribas2016} that the 
fluorescent effect of the central galaxy cannot explain the observed surface brightness profiles 
at distances $r\gsim 20$ physical kpc. 

   The non-linear clustering of objects derived from the hierarchical Cold Dark Matter model of 
structure formation predicts that a significant fraction of the faint sources likely reside around more 
massive, brighter galaxies. Therefore, star-forming regions and galaxies surrounding the central galaxy 
(satellite sources) may provide additional contributions to the extended halos at large distances from 
the center, $r\gsim30$ pkpc, via the nebular radiation produced `in-situ' in their ISM, and inducing fluorescent emission in the CGM of the central galaxy \citep[e.g.,][see \citealt{Maiolino2017} for a 
recent detection of star formation within outflows]{Shimizu2010,Matsuda2012,Lake2015,Momose2016}. 
Although most of the satellites are probably too faint to be resolved individually, their overall 
collective emission may be detectable, similarly  to the method of intensity mapping on large scales 
\citep[e.g.,][]{Chang2010,Visbal2010,Carilli2011,Gong2011,Silva2013,Dore2014,Pullen2014,Croft2016,
Li2016}. We addressed the relevance of satellite sources in \cite{Masribas2016}, accounting for the 
clustering of ionizing radiation which, in turn, yields to enhanced fluorescent \lya emission. Our results 
demonstrated that fluorescence alone cannot explain the observed profiles but its contribution can be 
up to $\sim50\%$ out to $r \sim 30$ pkpc if conditions of high escape fraction of ionizing photons and 
cold gas covering factor are accomplished.

   In the present work, we focus on the nebular emission (`in-situ' production) from the satellite 
sources. This analysis is important because, as we will demonstrate, we are able to reproduce 
the observed \lya and UV surface brightness profiles, which supports the notion that 
faint satellite sources can explain the extended LAHs. We self-consistently also predict H$\alpha$ 
and continuum surface brightness profiles for different models 
and parameters, which will be testable with future JWST observations. We show how the 
observations of H$\alpha$ surface brightness profiles will serve to clearly distinguish between the 
mechanisms that give rise to spatially extended emission, and will place constraints on halo star 
formation, in addition to the current UV measurements. 

   Obtaining tighter constraints to the presence of radiation sources in the halo of more massive galaxies 
allows for assessing the important role that faint objects played in the total cosmic photon budget 
\citep[see, e.g.,][]{Nestor2011,Nestor2013,Alavi2014,Garel2016} and, for high enough redshifts, 
their contribution to the reionization of the Universe \citep{Kuhlen2012,Robertson2013}.
Interestingly, \cite{Croft2016} recently reported an excess of \lya emission resulting from their 
cross-correlation between \lya surface brightness and quasars from the Sloan Digital Sky Survey III 
\citep[SDSS-III;][]{Eisenstein2011} Baryon Oscillation Spectroscopic Survey \citep[BOSS;][]{Dawson2013}. 
\cite{Croft2016} argue that, if their measured \lya emission is driven by star formation, this 
results in a star formation rate density $\sim30$ times larger than what is obtained from LAE surveys,  
although consistent with dust-corrected UV continuum analyses. 
The star formation scenario, however, needs to invoke an escape fraction for \lya $\sim100\%$, and 
strong radiative transfer effects. Our work can be viewed as a complementary experiment at 
smaller scales, where we `cross-correlate' deeper \lya intensity images with LAEs.

   We perform calculations considering the spatially extended emission observed around 
Lyman Alpha Emitters (LAEs) at redshift $z=3.1$, which allows for a comparison with the 
results by \cite{Momose2014} and \cite{Matsuda2012}. Our paper is structured as follows: 
In \S~\ref{sec:formalism}, we detail the formalism and adopted values for the parameters 
in the calculation of the surface brightness profiles for the continua, H$\alpha$ and Ly$\alpha$. 
We present the results for several models in \S~\ref{sec:sb}, and provide a discussion in 
\S~\ref{sec:discussion}, before concluding in \S~\ref{sec:conclusion}. Appendix \ref{sec:distrib} 
addresses the implications of the luminosity function parameter values, in terms of spatial and 
luminosity distribution of satellite sources around the central galaxy. In Appendix \ref{sec:snr},  
we detail the calculations of the signal-to-noise ratio for our predicted observations with JWST.

   We assume a flat $\Lambda$CDM cosmology with values ${\rm \Omega_{\Lambda}=0.7}$, 
${\rm \Omega_{ m}}=0.3$ and ${\rm H_0=68\, km\,s^{-1}\,Mpc^{-1}}$.

\section{Formalism}\label{sec:formalism}

   We present a simple analytic formalism that works with integrated 
properties of the entire emitting population, which allows to circumvent the 
modelling of individual sources when calculating the surface brightness profiles. 

   We demonstrated in Mas-Ribas \& Dijkstra (2016) that the fluorescent radiation 
from a central galaxy with ${\rm SFR\sim10\,M_{\odot}\,yr^{-1}}$ only dominates at distances 
$\lsim 20-30$ pkpc from the center, and at a level that strongly depends on the characteristics 
of the circumgalactic gas. In addition, at such small distances, the profile of the central galaxy is 
significantly driven by the point-spread function (PSF) of the instrument \citep{Momose2014}. 
Owing to these uncertainties, we here ignore the central galaxy and limit our calculations to 
distances $> 10$ pkpc. 

   We use a similar formalism to that applied in \cite{Masribas2016}, to which we 
refer the reader for details. Briefly, the \lya and H$\alpha$ surface brightness at impact 
parameter $b$ equals 
\begin{equation}\label{eq:sb}
SB_{x}(b) = \frac{2}{(1+z)^{4}}\int_b^{R_{max}^{\alpha}} \bar{\epsilon}^{\rm sat}_{x} [1+\xi_x(r)] f_{\rm esc}^x\frac{r{\rm d}r }{\sqrt{r^2-b^2}} ~,
\end{equation} where `$x$' stands for \lya or H$\alpha$. The factor ${(1+z)^{-4}}$ accounts 
for the surface brightness dimming.  The factor $\bar{\epsilon}^{\rm sat}_{x}$ denotes
the integrated volume emissivity in satellite galaxies (see \S~\ref{sec:emiss}), the term 
$[1+\xi_x(r)]$ denotes the boost in emissivity due to clustering of sources around the central 
galaxy (see \S~\ref{sec:xi}), and $f_{\rm esc}^x$ denotes the escape fraction (see 
\S~\ref{sec:fesc}). Finally, the value for the upper limit of the integral extends to infinity for 
the Abel transformation used above but we limit its value accounting for the line-shift due to 
the expansion of the universe as
\begin{equation}
R_{\rm max}^{\alpha}=\frac{1}{2} \frac{c}{H(z)}  \frac{{\rm d}\nu_{\alpha}}{{\nu_{\alpha}(z)}}~.
\end{equation}
$H(z)$ denotes the Hubble parameter at a given redshift, $c$ is the speed of light and 
d$\nu_{\alpha}/\nu_{\alpha}=0.02$ accounts for the line-shift for apertures in 
narrowband surveys of $\sim100\,{\rm \AA}$, e.g., \cite{Matsuda2012}. This approach implies  
$R_{\rm max}^{\alpha}\sim3$ pMpc, but we have tested that our results show only differences  
of a factor $\sim2$ at large distances, $r\gsim80-100$ pkpc, when setting the upper limit 
within the range $300\, {\rm pkpc}<R_{\rm max}^{\alpha}<5$ pMpc. 

   We calculate the UV surface brightness at $1500\,\angs$ rest-frame as 
\begin{equation}\label{eq:sbuv}
SB_{\rm UV}(b) = \frac{2}{(1+z)^{3}}\int_b^{R_{max}^{\rm UV}} \bar{\epsilon}^{\rm sat}_{\rm UV} [1+\xi_{\rm UV}(r)] f_{\rm esc}^{\rm UV} \frac{r{\rm d}r}{\sqrt{r^2-b^2}} ~,
\end{equation}
where we use the parameters for UV radiation, and have multiplied Eq.~\ref{eq:sb} by  
$(1+z)$, since the UV surface brightness is measured as a flux {\it density} (in units of 
inverse frequency) per unit solid angle. 

We compute the surface brightness for the visible continuum (VIS) as 
\begin{equation}\label{eq:vis}
SB_{\rm VIS}(b)= \frac{(1+z)}{\rm EW_{H\alpha}}\frac{\lambda_{\rm H{\alpha}}^2}{c}\, SB_{\rm H\alpha}(b)   ~.
\end{equation}
We derive the VIS emission using the \ha equivalent width because visible radiation 
is not commonly used as a star formation estimator, therefore not providing a relation between 
star formation and luminosity at a specific wavelength, unlike Ly$\alpha$, \ha and UV in 
Eq.~\ref{eq:phi} \citep[see][for a complete review]{Kennicutt2012}. We assume a flat spectrum around 
\ha and a line equivalent width ${\rm EW_{H\alpha}=300\,\angs}$ \citep[rest-frame; e.g.,][and 
references therein]{Queralto2016}. In Eq.~\ref{eq:vis}, $\lambda_{\rm H\alpha}$ and $c$ represent 
the \ha wavelength at rest and the speed of light, respectively, applied to obtain the surface brightness 
in units of inverse frequency. For completeness, we will also explore the ranges $450\geq {\rm 
EW_{H\alpha}\,[\AA] \geq 150}$ and $700\geq {\rm EW_{H\alpha}\,[\AA] \geq 50}$.

\subsection{Volume Emissivity, $\bar{\epsilon}_{x}^{\rm sat}$}\label{sec:emiss}

   The integrated volumetric emissivity (i.e., volume emissivity) in faint satellites is given by   
\begin{equation}\label{eq:sfrlum}
\epsilon^{\rm sat}_{x} = C_{x}\rho^{\rm sat}_{\rm SFR} ~,
\end{equation} 
where $\rho^{\rm sat}_{\rm SFR}$ denotes {\it the star formation rate density in faint satellites}. 
We are interested in the contribution to the star formation rate density from sources fainter than 
$M_{\rm UV}\equiv M^{\rm sat}_{\rm UV}=-17$, which corresponds roughly to the minimum UV 
luminosity of unlensed galaxies that can be detected directly \citep[e.g.,][]{Bouwens2015,
Finkelstein2015}. For a UV luminosity function 
with faint-end slope $\alpha=-1.7\, (-1.5)$, this approach translates to extrapolating the LF 
to $M_{\rm UV}\sim-12\, (M_{\rm UV}\sim-10)$ \citep[][]{Kuhlen2012,Alavi2016,Lapi2016,Livermore2017}. 
The integrated cosmic star formation 
rate density in the {\it observed population} of star forming galaxies is $\rho_{\rm SFR}\sim 0.1\,
{\rm M_{\odot}\,yr^{-1}\,cMpc^{-3}}$ at $z\sim3$ \citep[see, e.g.,][]{Hopkins2006,Bouwens2015,
Khaire2015b,Robertson2015}. We assume that $\rho^{\rm sat}_{\rm SFR}=\rho_{\rm SFR}$, for 
simplicity. This assumption depends in detail on the faint-end slope of the UV luminosity function 
(LF) at $M_{\rm UV} > M^{\rm sat}_{\rm UV}$, on $M^{\rm sat}_{\rm UV}$ itself, and the UV 
magnitude down to which we integrate this LF. The precise value for  $\rho^{\rm sat}_{\rm SFR}$ 
is, therefore, highly uncertain, and our results scale linearly with the value for this parameter. 
The constant $C_{x}$ represents the standard conversion factor from SFR into UV luminosity 
density, H$\alpha$ and Ly$\alpha$ luminosities, and is given by
\begin{equation}
C_x= \left\{ \begin{array}{ll} 
1.30\times 10^{42}\hspace{1mm}\frac{{\rm erg}\hspace{1mm} {\rm yr}}{{\rm s}\hspace{1mm}M_{\odot}}& (\mbox{Ly}\alpha)~;\\  
1.26\times 10^{41}\hspace{1mm}\frac{{\rm erg}\hspace{1mm} {\rm yr}}{{\rm s}\hspace{1mm}M_{\odot}} &(\mbox{H}\alpha)~;\\
8.00\times 10^{27}\hspace{1mm}\frac{{\rm erg}\hspace{1mm} {\rm yr}}{{\rm s}\hspace{1mm}M_{\odot}\hspace{1mm}{\rm Hz}} & (\mbox{UV})~.\end{array}  \right.
\label{eq:phi}
\end{equation} 
The conversion factor for the UV continuum comes from \citet{Madau1998}, and 
for H$\alpha$ from \citet{Kennicutt1998}. We obtain the conversion factor for Ly$\alpha$ 
from H$\alpha$, assuming the common $L_{\rm Ly\alpha} = 8.7 L_{\rm H\alpha}$ ratio 
\citep{Brocklehurst1971,Barnes2014,Dijkstra2014}, which assumes case-B recombination. 
We caution that these conversion factors, especially for Ly$\alpha$, can vary depending on the 
metallicity, initial mass function (IMF), and ages of the stellar population \citep[][]{Raiter2010,
Masribas2016b}. Additionally, the value of $C_{\rm Ly\alpha}$ strongly depends on the  
\lya rest-frame equivalent width of the sources. We will demonstrate in \S~\ref{sec:cx} that accounting 
for this dependence over the faint satellite population has a significant impact on the results.

\subsection{Clustering of Emission, $[1+\xi_x(r)]$}\label{sec:xi}

   The cross-correlation function of emission around LAEs is proportional to the matter density 
field and can be written as $\xi_{x}(r)=b_{x}(r)b_{\rm LAE}(r)\xi(r)$. The term $\xi(r)$ denotes 
the non-linear dark matter correlation function obtained using CAMB \citep{Lewis1999}. 
The terms $b_{\rm LAE}(r)$ and $b_x(r)$ are the distance-dependent LAE and emission biases, 
respectively. We discuss these terms below. 

\subsubsection{The ${\rm LAE}$ bias, $b_{\rm LAE}(r)$}\label{sec:blae}

\begin{figure}
\includegraphics[width=0.49\textwidth]{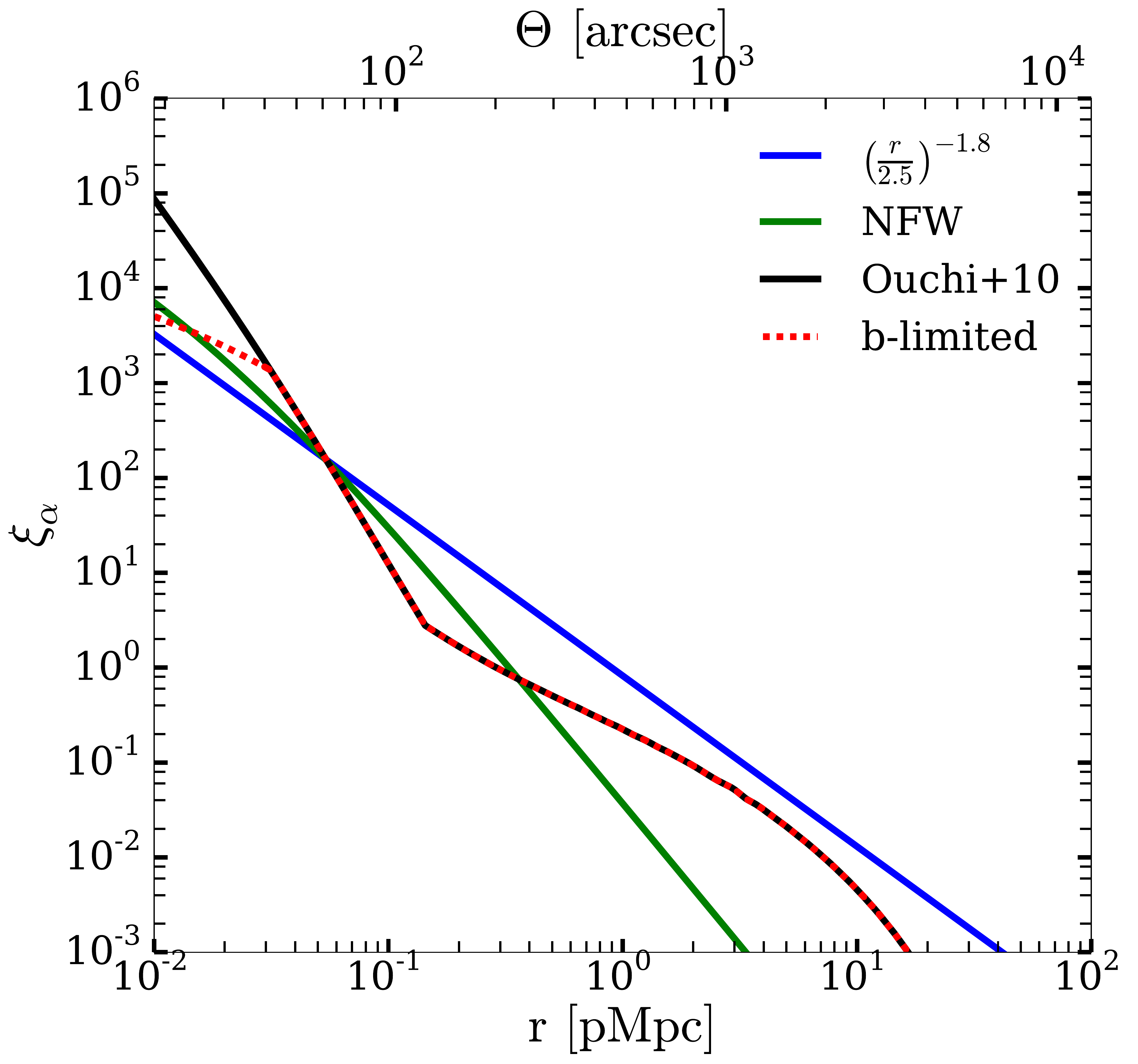}
\caption{\lya emission correlation functions as a function of distance from the central 
galaxy for different clustering models. The {\it solid black line} denotes our fiducial model, 
derived by extrapolating observational constraints on the scale-dependent bias by 
\cite{Ouchi2010}. The {\it dashed red curve} displays the same model, but with the bias 
limited to a value 10 (the maximum value inferred by Ouchi et al. 2010) . The {\it blue line} 
indicates the common power-law clustering of LAEs at redshift $z\sim 3$. The {\it green line} 
denotes the NFW profile (see text).} 
\label{fig:corr}
\end{figure}

   Following \citealt{Masribas2016} (see their Appendix B for details) our fiducial 
model ({\it solid black line} in Figure \ref{fig:corr}) adopts $b_{\rm LAE}(r)$ based on 
observations by \cite{Ouchi2010}, who measured $b_{\rm LAE}(r)$ to increase to 
$b_{\rm LAE}(r)\sim 10$ down to $r \sim 20$ pkpc. We tested in our previous work that 
using this clustering we obtained an overdensity $\delta_{\rm LAE}\sim1.5$, averaged 
over a radial distance of $2\,{\rm Mpc\,h^{-1}}$ from the central galaxy, consistent with 
the values reported by \cite{Matsuda2012}. To quantify how much our results depend 
on extrapolating $b_{\rm LAE}(r)$ down to smaller scales, we have also repeated our 
calculations, but limiting the $b_{\rm LAE}(r)$ to a maximum value of 10. This model 
is represented in Figure \ref{fig:corr} as the {\it dotted red line}. In addition, the 
observational uncertainties reported by \cite{Ouchi2010} for the bias at $r \sim 20$ kpc are of 
the order $\sim 50\%$, consistent at a $2\sigma$ level with the bias obtained assuming a 
power-law correlation function. Owing to these large uncertainties for the bias at small 
scales, we also explore other clustering prescriptions in \S~\ref{sec:otherclust} below.

\subsubsection{Emission bias, $b_x(r)$}\label{sec:bias}

   The term $b_{\alpha}(r)$ expresses the distance-dependent bias of the Ly$\alpha$ emission, 
which we assume to differ from that of LAEs by a constant, i.e.,  $b_{\alpha}(r)=k\,
b_{\rm LAE}(r)$. The bias $b_{\alpha}(r)$ represents the Ly$\alpha$ 
luminosity-weighted average of the entire satellite population. Its value thus depends on the 
faint-end slope of the Ly$\alpha$ luminosity function \citep[$\alpha_{{\rm Ly}\alpha}$, which is 
likely steeper than the UV-LF; see, e.g.,][]{Gronke2015b,Konno2016}, although, as long as 
$\alpha_{{\rm Ly}\alpha} > -2$,  we expect that the bias is set by the most luminous satellites 
with $M_{\rm UV}\sim M_{\rm UV}^{\rm sat} = -17$. \cite{Gronke2015b} have shown that 
observational constraints on $M_{\rm UV}$-dependent Ly$\alpha$ equivalent width (EW) PDFs 
imply that the faintest LAEs ($L_{\rm Ly\alpha}\sim 10^{42}$ erg s$^{-1}$) are associated with 
galaxies with $M_{\rm UV}\sim -17.5$ (see their Figure~3). This result suggests that the UV-brightest 
satellites may cluster like LAEs, with $k\sim 1$, although fainter sources might present 
values $k>1$. \citet{Croft2016} argue that $b_{\alpha}(r)$ might be further boosted by radiative 
transfer effects due to the resonant nature of the \lya radiation \citep[e.g.,][]{Zheng2011b}. 
To be conservative, we adopt $k=1$ in our fiducial model\footnote{It is worth pointing out that 
the value of $k$ formally cannot be chosen independently of the escape fraction (see 
\S~\ref{sec:fesc}); a high $k$ value implies that radiative transfer in the CGM/IGM is important. 
In order to reproduce the observed Ly$\alpha$ LFs of LAEs, one then requires that 
$f_{\rm esc}^{\rm Ly\alpha}\sim 1$ \citep[see][]{Zheng2010}.}.

   Our fiducial model assumes that for both, UV and H$\alpha$ emission, we have 
$b_{\rm UV}(r)=b_{\rm H\alpha}(r)=b_{\alpha}(r)=b_{\rm LAE}(r)$, i.e $k=1$. This 
choice is motivated by the 
discussion above, while noticing that radiative transfer cannot further enhance $k$ in these 
cases. Our predicted surface brightness profiles again scale linearly with $k$.

\subsubsection{Alternative clustering prescriptions}\label{sec:otherclust}

   We consider two alternative clustering prescriptions:
\begin{enumerate}[leftmargin=*]
\item The distribution of satellites follows that of dark matter in a Navarro-Frenk-White (NFW) 
profile \citep{Navarro1997}, normalized to be the same as the other clustering estimators 
at $r \sim 60$ pkpc (similar to the value of the virial radius for the central galaxy). This model 
is represented by the {\it green line} in Figure \ref{fig:corr}. The density profile in the NFW 
model equals
\begin{equation}
\rho (r) = \frac{\delta_c \rho_c(z)}{r/r_s (1 + r/r_s)^2} ~,
\end{equation}
where $\rho_c(z)$ is the critical density of the universe at redshift $z$, $\delta_c$ is 
the characteristic overdensity and $r_s$ is the scale radius of the dark matter halo.
The overdensity $\delta_c$ can be expressed as 
\begin{equation}
\delta_c = \frac{\Delta}{3}\frac{c_{\rm NFW}^3}{\log(1+c_{\rm NFW}) - c_{\rm NFW}/(1+c_{\rm NFW})}~, 
\end{equation}
where $\Delta=18\pi^2$ is the density contrast from \cite{Bryan1998} and $c_{\rm NFW}=4$ 
is the concentration parameter at $z=3.03$ from \cite{Zhao2009}. We obtain the 
scale radius from the expression $c_{\rm NFW} \equiv r_h/r_s$, where $r_h$ is the 
halo virial radius, computed as
\begin{equation}
r_h=\left(\frac{3M_h}{4\pi \Delta \rho_c(z)}\right)^{1/3}~.
\end{equation}
This expression emerges from considering that the mean density of the halos within the 
virial radius is $\Delta \rho_c(z)$ \citep{Sadoun2016}. We have assumed an 
LAE halo mass $\log M_h=11.5\,{\rm M_\odot}$, consistent with the observed range of 
LAE masses in \cite{Ouchi2010}.

\item We extrapolate the common LAE power-law two-point correlation function, with scale 
length $r_0=2.5\,{\rm Mpc\,h^{-1}}$ and power-law index $\alpha_c=-1.8$ 
\citep[e.g.,][]{Gawiser2007, Kovac2007, Ouchi2003, Ouchi2010, Guaita2010,Bielby2015} 
down to small scales. This clustering profile is denoted by the {\it blue solid line} 
in Figure \ref{fig:corr}. The power-law function presents differences with our fiducial model at 
distances above $\sim 60$ pkpc and at tens of pkpc from the center. In this last region is 
where the non-linear clustering effects, not captured by the power-law, are important, 
therefore higher values for the fiducial function are  expected.
\end{enumerate}

\begin{figure*}
\includegraphics[width=0.48\textwidth]{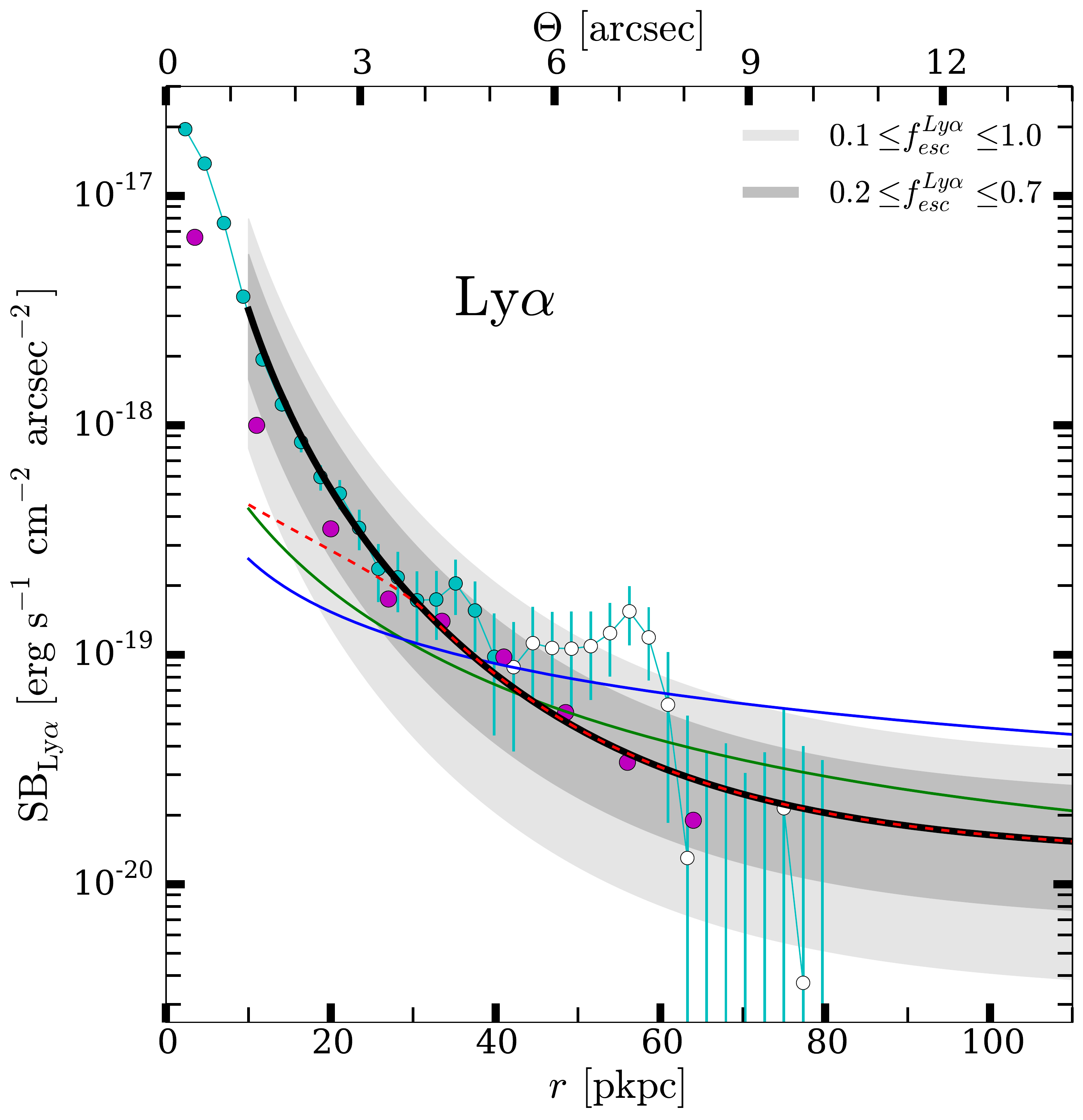}\includegraphics[width=0.48\textwidth]{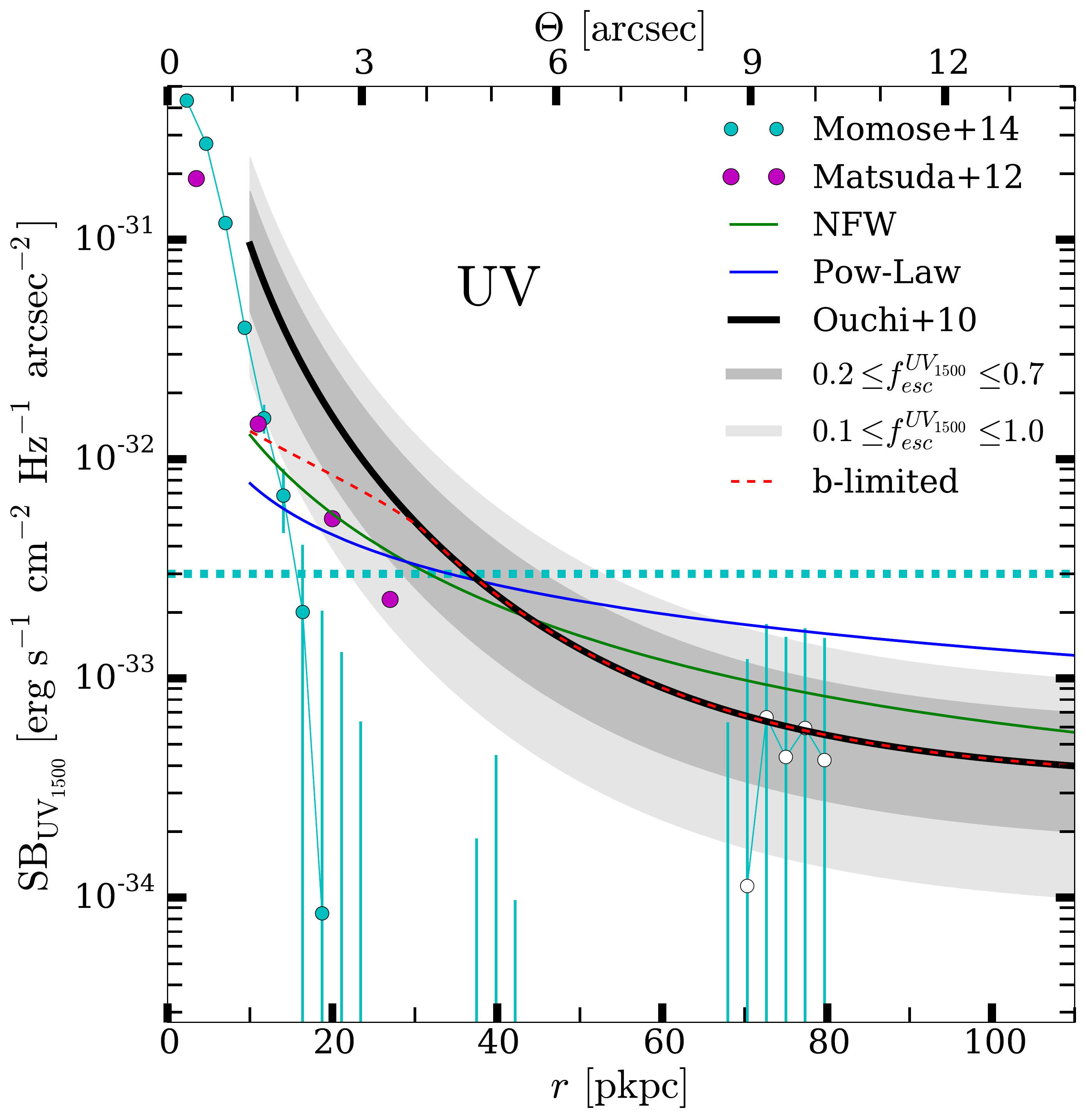}
\caption{{\it Left panel:} Radial \lya surface brightness profiles with physical distance from the 
central galaxy for different models. Lines and colors refer to the same models as in Figure 
\ref{fig:corr}. {\it Magenta points} denote an estimation of the data for the LAE overdensity range 
$2.5<\delta_{\rm LAE}<5.5$ in \cite{Matsuda2012}. The {\it cyan dots and bars} represent the 
mean values and uncertainties from the observations at $z=3.1$ by \cite{Momose2014}, 
respectively. {\it White dots} indicate the regions where the measurements are not reliable due 
to systematic effects. The fiducial model is denoted by the {\it solid black line} considering 
$f_{\rm esc}^{\rm Ly\alpha}=0.4$, and the two {\it shaded areas} display the regions 
$0.2\le f_{\rm esc}^{\rm Ly\alpha}\le 0.7$ and $0.1\le f_{\rm esc}^{\rm Ly\alpha}\le 1.0$ for the 
same model. {\it Right panel:} UV surface brightness profile. Lines and symbols are the same 
as in the left panel. The {\it horizontal dashed cyan line} indicates the region below which 
systematic effects in the observational data by \cite{Momose2014} are important. The two 
{\it shaded areas} display the regions within the same $f_{\rm esc}^{\rm UV}$ ranges as for \lya 
around the fiducial model. } 
\label{fig:sb}
\end{figure*}

\subsection{Escape Fraction, $f_{\rm esc}^{x}$}\label{sec:fesc}

   The \lya escape fraction, $f_{\rm esc}^{\rm Ly\alpha}$, has been constrained observationally 
to be $f_{\rm esc}^{\rm Ly\alpha}\sim 20\%$ at $z\sim 3$ from the \lya and UV luminosity 
functions  \citep[e.g.][]{Blanc2011, Hayes2011b}, and Ly$\alpha$ and star formation analysis  
\citep{Dijkstra2013}. However, we caution that all the observations have constrained the 
`effective' escape fraction, which denotes the fraction of Ly$\alpha$ photons that reaches the 
observer. As mentioned previously, in some models {\it all} Ly$\alpha$ photons escape from the 
ISM, but then scatter in the CGM/IGM to form halos (in these same models Ly$\alpha$ radiative 
transfer causes $k > 1$). These photons would not have been considered in traditional 
measurements of Ly$\alpha$ luminosity functions \citep[up to a fraction $40\% - \gsim 90\%$ of the 
total \lya flux, as argued by][see also \citealt{Drake2016}]
{Wisotzki2015} and, therefore, not considered for current observational constraints on 
$f_{\rm esc}^{\rm Ly\alpha}$. Also, there is observational evidence that the Ly$\alpha$ escape 
fraction increases towards lower UV-luminosities \citep[e.g.,][see also \citealt{Dijkstra2016} and 
references therein]{Japelj2016}. While observations find $f_{\rm esc}^{\rm 
Ly\alpha}\sim20\%$, we consider this value likely a lower-limit and adopt the range $0.1\le f_{\rm 
esc}^{{\rm Ly}\alpha}\le 1.0$ throughout, with a fiducial value $f_{\rm esc}^{\rm Ly\alpha}=40\%$.

   We adopt the same range and conservative fiducial value for UV and 
H$\alpha$ escape fractions. We generally expect that $f_{\rm esc}^{\rm UV} \geq 
f_{\rm esc}^{\rm Ly\alpha}$ because UV photons are not affected by radiative transfer effects, 
i.e., resonant scattering that increases the chance to be destroyed by dust \citep[see, e.g., Figure 
7 in ][]{Garel2015}. The escape fraction 
of H$\alpha$ can be even larger than that of UV, due to the wavelength dependence of 
the dust extinction curve \citep[see, e.g.,][]{Pei1992,Calzetti1994,Calzetti2000,Gordon2003}.

\section{Surface brightness profiles}\label{sec:sb}

   We present the resulting surface brightness profiles below. It is important to keep in mind  
that these results are degenerate in the product of emissivity, escape fraction and bias, 
$\bar{\epsilon}^{\rm sat}_x\, f_{\rm esc}^x\, b_x$, where $x$ refers to UV, H$\alpha$ and 
Ly$\alpha$.   

\subsection{\rm \lya}
   The {\it left panel} in Figure \ref{fig:sb} shows the predicted \lya surface brightness profile at 
$r>10$ pkpc. The {\it black solid line} denotes the fiducial model, and the {\it shaded areas} 
indicate the range of surface brightness profiles we get by varying $0.2\le f_{\rm esc}^{\rm 
Ly\alpha}\le 0.7$ ({\it dark}) and $0.1\le f_{\rm esc}^{\rm Ly\alpha}\le 1.0$ ({\it light}). These 
ranges give an idea of the effect of a possible radial variation of the escape fraction due to the 
decrease of neutral gas with distance. The 
{\it blue}, {\it green} and {\it dotted red lines} represent the power-law, NFW and `bias-limited' 
models, respectively (for our fiducial choice $f_{\rm esc}^{\rm Ly\alpha}=0.4$). The {\it light 
blue dots} represent the data and uncertainties from the observations by \cite{Momose2014} 
at $z=3.1$, which are not reliable at $r>40$ pkpc due to systematics \citep[and therefore 
represented with {\it open circles};][see also \citealt{Feldmeier2013}]{Momose2014}. 
{\it Magenta dots} represent the data in the LAE overdensity bin $2.5<\delta_{\rm LAE}<5.5$ by  
\cite{Matsuda2012}, which we also used in \cite{Masribas2016} given the value of our 
LAE overdensity.  

   Our fiducial model reproduces the observations well within the range $20\lsim r \lsim 40$ pkpc. 
At shorter distances, the Ly$\alpha$ surface brightness may be enhanced by resonantly scattered 
\lya that escapes from the central LAE and/or by fluorescence \citep{Masribas2016}. Systematics 
may in turn affect the data at $r \gsim 40$ pkpc, although the fiducial model reproduces the data 
from \cite{Matsuda2012} at these scales remarkably well. The other clustering prescriptions 
reproduce the observed surface brightness levels to within a factor of $\sim 2$ at  $20\lsim r 
\lsim 40$ pkpc. In general, they give rise to flatter surface brightness profiles, which reflects that 
in these models $\xi_{\alpha}$ is flatter at $r \lsim 100$ pkpc. The impact of the different clustering 
prescriptions becomes more severe at $r\lsim20$ pkpc. However, as we mentioned previously, 
here we expect the surface brightness profile to be enhanced by Ly$\alpha$ and LyC photons 
that escaped from the central LAE.

\begin{figure*} 
\includegraphics[width=0.48\textwidth]{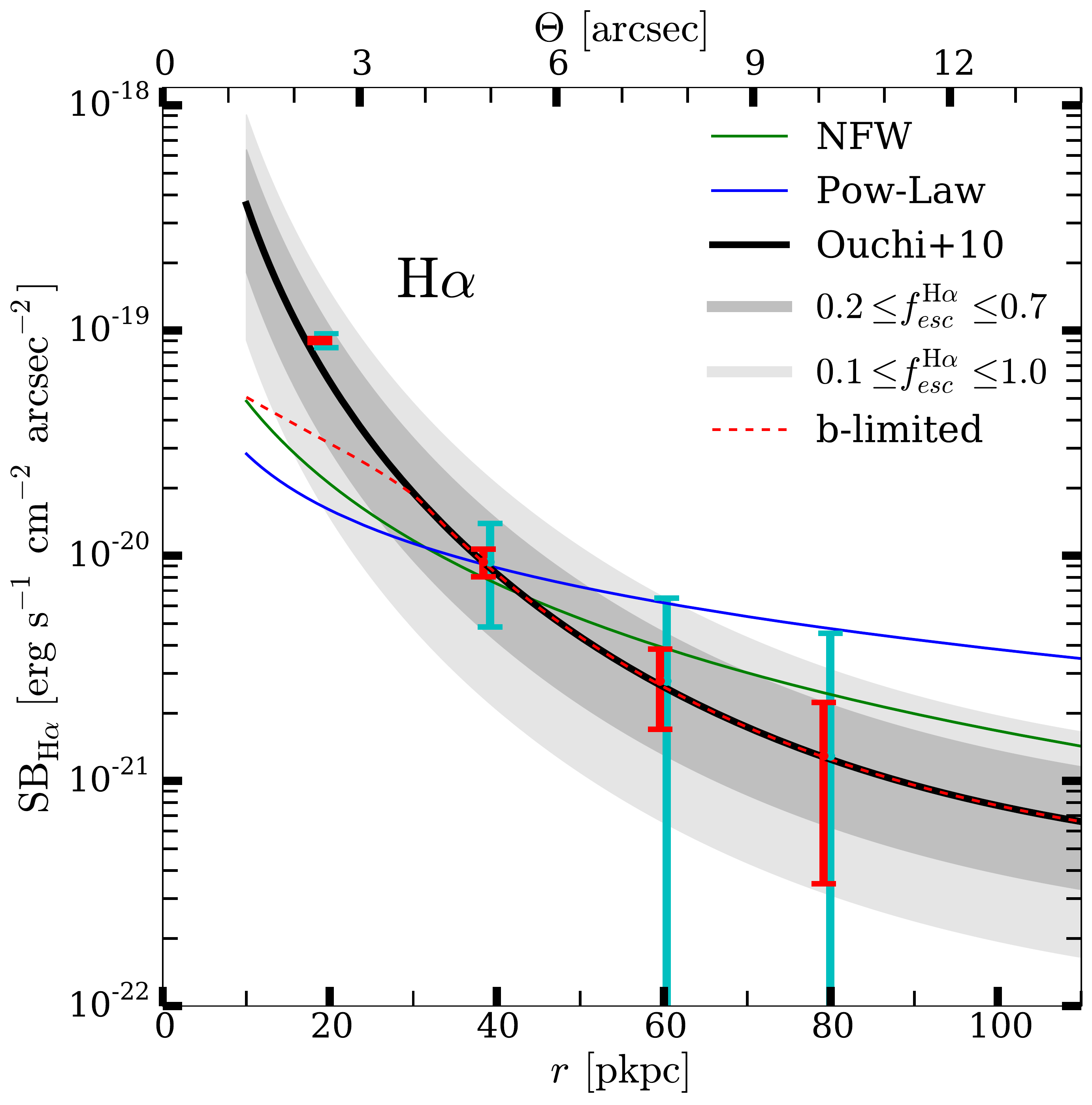}\includegraphics[width=0.48\textwidth]{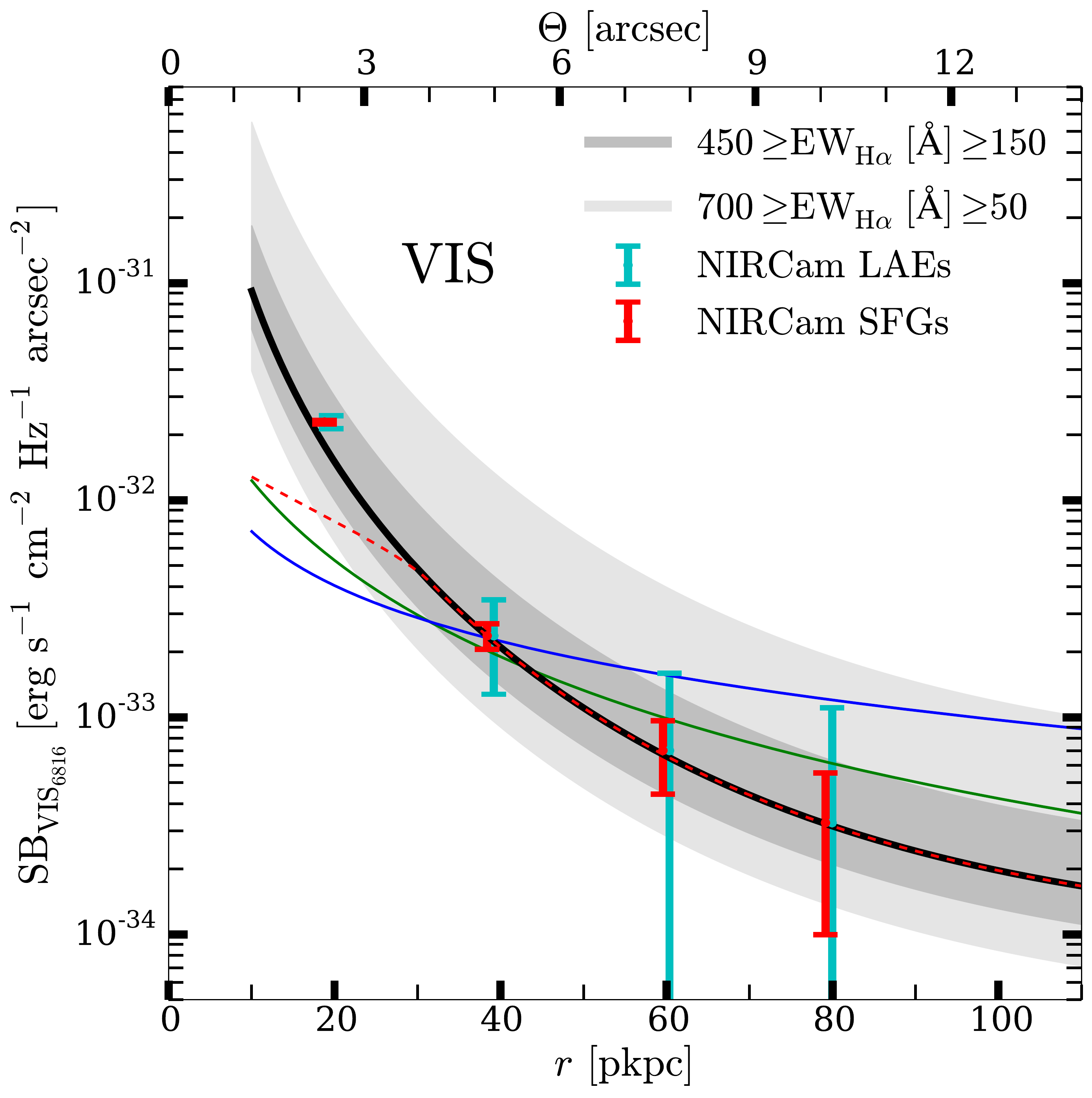}
\caption{{\it Left panel:} Radial H$\alpha$ surface brightness profiles predicted by our different models. 
Lines and colors are the same as in Figure \ref{fig:sb}. The two {\it shaded areas} display the regions 
$0.2\le f_{\rm esc}^{\rm H\alpha}\le 0.7$ and  $0.1\le f_{\rm esc}^{\rm H\alpha}\le 0.1$ for the fiducial 
model ({\it black solid line}). The {\it blue} ({\it red}) {\it error bars} indicate the values and uncertainties 
for the predicted observations of halos around LAEs (SFGs) described in \S~\ref{sec:obs} 
(\S~\ref{sec:obs2}). {\it Right panel:} Same as in the left panel but considering the visible continuum 
emission at ${\rm 6816\,\angs}$ rest-frame, computed assuming ${\rm EW_{H\alpha}=300\,\angs}$. 
The shaded areas represent the ranges $450\geq {\rm EW_{H\alpha}\,[\AA] \geq 150}$ and $700\geq 
{\rm EW_{H\alpha}\,[\AA] \geq 50}$ for the fiducial model. The red and blue data points at $r=20$ 
pkpc fall slightly above the solid black line due to averaging the signal around such a steep regions. 
The red vertical lines have been slightly shifted from their original position to facilitate the visualization.} 
\label{fig:sbha}
\end{figure*} 

\subsection{\rm UV}

   The {\it right panel} in Figure \ref{fig:sb} shows the predicted UV surface brightness profiles. 
We use the same symbols and colors as in the {\it left panel}. The {\it horizontal dashed line} 
shows the UV surface brightness level below which the data by \cite{Momose2014} is affected 
by systematics. This figure shows that at $20\lsim r \lsim 40$ pkpc our fiducial model predicts a 
UV profile above the observations by a factor of $\sim 3$, while the other clustering prescriptions 
lie within a factor of $\sim 1.5 - 2$. Our fiducial model thus results in an excess - by a factor of $\sim 3$ 
- of UV emission in the halos of LAEs. This result may reflect an overestimated star formation rate 
density in faint galaxies (i.e., $\rho^{\rm sat}_{\rm SFR}$). Based on analysis and modeling of 
Hubble Space Telescope observations aimed at detecting long-duration gamma-ray bursts host 
galaxies at high redshift, \cite{Trenti2012} inferred that $\sim 30 \%$ ($\sim 40 \%$) of the total star 
formation at $z\sim3$ ($z \lsim 5$) occurs in galaxies too faint to be directly detected. This result is 
broadly consistent with the difference observed here, although the reduced emissivity value would 
also result in \lya profiles below the observations by the same factor if no other parameters are tuned. 
An overestimated escape fraction $f_{\rm esc}^{\rm UV}$ would produce the same effect, though we 
consider this possibility unlikely.  Alternatively, we may have overestimated the abundance of sources 
in the halo of LAEs due to clustering. We investigate the predicted luminosity and spatial distributions 
of satellites for various models, and the dependence on luminosity function parameter values in 
Appendix \ref{sec:distrib}. In \S~\ref{sec:cx}, we further discuss the significant effect of a likely 
evolution of the \lya rest-frame equivalent with the UV magnitude of the sources.

\subsection{\rm H$\alpha$}

   Our predictions can be tested with future observations of \ha surface brightness 
profiles since \ha falls into the wavelength range covered by the James Webb Space Telescope 
\citep[JWST;][]{Gardner2006}. In addition, H$\alpha$ does not resonantly scatter, which simplifies 
interpreting its surface brightness profile compared to Ly$\alpha$, and enables distinguishing 
between the different possible origins of LAHs. 

   The {\it left panel} in Figure \ref{fig:sbha} displays the predicted H$\alpha$ surface brightness profiles. 
Our fiducial model is represented by the {\it solid black line} and (conservatively) assumes 
$f_{\rm esc}^{\rm H\alpha}=0.4$. The impact of varying $f_{\rm esc}^{\rm H\alpha}$ and other 
models are shown in the same way as in Figure \ref{fig:sb}. The fiducial profile rises above $10^{-19}\, 
{\rm erg\,s^{-1}\,cm^{-2}\,arcsec^{-2}}$ at distances $r\lsim 20$ pkpc. The {\it red} and {\it blue 
error bars} represent the predicted uncertainties on the surface brightness profile, as if it was observed 
by the near infrared camera (NIRCam) onboard JWST considering the two observational strategies 
described below. The surface brightness uncertainties are derived from the signal-to-noise ratio 
(SNR), which decreases from ${\rm SNR}\sim14$ ($\sim 48$) at $r=20$ pkpc to ${\rm SNR}\sim0.4$ 
($\sim 1.4$) at $r=80$ pkpc for halos around observed LAEs (star-forming galaxies, SFGs). We detail 
the calculations of the SNR in Appendix \ref{sec:snr}. The {\it left panel} in 
Figure \ref{fig:sbha} indicates that the H$\alpha$ emission predicted by the various 
models can be detected up to distances $r\gsim80$ pkpc when stacking the SFGs expected in the 
field of view (FOV)\footnote{\cite{Zhang2017} have already demonstrated in a recent work the 
strength of the stacking technic applied to \ha radiation around low redshift galaxies.}. Considering uniquely the emission around observed LAEs and our adopted 
observational strategy, NIRCam can prove the halos up to $r\sim40$ pkpc, yielding upper 
limits at larger distances (see below). However, the presence of star formation at 
large distances from the central LAEs can be assessed up to $r\sim80$ pkpc with 
observations of \ha and visible continuum radiation around star-forming galaxies ({\it red error bars}  
in both panels of Figure \ref{fig:sbha}.

\subsubsection{\rm NIRCam \ha observations of LAEs}\label{sec:obs}

   The Multi-object Spectroscopy\footnote{\url{https://jwst-docs.stsci.edu/display/JTI/NIRSpec+Multi+Object+Spectroscopy}} (MOS) observing mode of the near infrared spectrograph (NIRSpec) 
would be desirable for our observations, given the large FOV, high spectral resolution (up to 
$R\sim2700$), and the obtention of the spectra over a broad wavelength range. However, 
observations of nearby areas of the sky with contiguous (in the direction of dispersion) elements of 
the Micro-shutter Assembly (MSA) result in spectra overlapping. The Integral-field 
Unit\footnote{\url{https://jwst-docs.stsci.edu/display/JTI/NIRSpec+IFU+Spectroscopy}} 
(IFU) spectroscopy mode circumvents this problem with the use of 3-dimensional spectral imaging 
data cubes but, in this case, the FOV is smaller than the expected halo of a single galaxy (FOV$
\sim3"\times3"$). Owing to the impracticability of the above modes, we consider the imaging 
capabilities of NIRCam for our calculations.

   We adopt the narrow-band (NB) filter F323N, with a bandpass of $0.038\,\mu{\rm m}$, 
resulting in a resolution $R\sim85$. We use this filter because it is the one closer to the \ha 
wavelength of interest, but we note that it is centered at a wavelength $3.2\,\mu{\rm m}$, 
corresponding precisely to an \ha redshift  $z=3.9$. For this calculations, we assume the previous 
\ha flux and surface brightness at $z=3.1$, but we recalculate the sky background\footnote{For 
the calculation of the background, we adopt Eq.~22 in the NIRSpec technical note \url{http://www.stsci.edu/~tumlinso/nrs_sens_2852.pdf}.} at $z=3.9$, obtaining ${\rm SB_{sky}\, 
(3.20\,\mu{\rm m})=4\times 10^{-20}\,erg \,s^{-1}\,cm^{-2}\, \angs^{-1} \,arcsec^{-2}}$, consistent with 
the estimates by \cite{Giavalisco2002} for HST and the Spitzer/IRAC measurements by 
\cite{Krick2012}. We set the observing time to $10^4$ s. 

   We calculate the number of LAEs observable simultaneously in the FOV of NIRCam (FOV$= 2\times 
2'.2 \times 2'.2$) as follows: We integrate the LAE luminosity function at $z=3.1$ by \cite{Ouchi2008}, 
with the parameters in Table \ref{ta:sch}, for the luminosity range $10^{42} \le L_{\rm Ly\alpha}\,
({\rm erg\,s^{-1}}) \le 10^{44}$. This calculation yields a space density of LAEs $n_{\rm LAE}
\sim2\times 10^{-3}\,{\rm cMpc^{-3}}$, in agreement with the findings by \cite{Ciardullo2006}. 
The selected filter results in a redshift depth $\Delta z = 0.058$ centered at $z=3.9$, giving rise to 
the simultaneous observation of $\sim 7$ LAEs in the FOV. 

   Considering LAEs with luminosities $L_{\rm LAE}>10^{42}\,{\rm erg\,s^{-1}}$, we can prove LAHs 
up to distances $\sim 40$ pkpc, covering entirely the range of radii out to which the extended 
emission has been detected around LAEs. However, LAEs with these luminosities account for a 
small fraction of the 
total star-forming galaxy population. JWST surveys, as those already proposed by the 
NIRSpec and NIRCam GTO teams in the GOODS and CANDELS fields\footnote{\url{https://confluence.stsci.edu/display/STUCP/JWST+Guaranteed+Time+Observers+Cycle+1+Plans}}, 
will detect a larger number of star-forming galaxies by means of the continuum and \ha radiation. 
We show below that stacking a larger sample of galaxies will enable proving extended \ha emission 
at larger distances from the center of galaxies, and reaching low surface brightness levels, useful for 
assessing the role of cooling radiation. 

\subsubsection{\rm NIRCam \ha observations of SFGs}\label{sec:obs2}

   We predict the extended \ha emission around star-forming galaxies using the same observing 
configuration as above, but we estimate the number 
of SFGs as follows: We integrate the UV luminosity function with the parameters 
by \cite{Kuhlen2012} listed in Table \ref{ta:sch} within the range $-24\le {M_{\rm UV}}\lsim -17$,  
resulting in a space density $n_{\rm SFG}\sim 2\times10^{-2}\,{\rm cMpc^{-3}}$. The upper limit, 
$M_{\rm UV}\sim -17$, rises from considering $L_{\rm UV_{1500}}^{\rm min}\sim 0.025\, L_{\rm 
UV}^*$, and is consistent with the current limit of (unlensed) galaxy surveys \citep{Finkelstein2015}. 
The obtained space density results in the simultaneous observation of $\sim 86$ SFGs in the FOV. 

   We refer the reader to Appendix \ref{sec:snr} for a detailed description of the signal-to-noise 
ratio calculations for the two above strategies.

\subsection{\rm VIS}\label{sec:vis}

   The {\it right panel} in Figure \ref{fig:sbha} shows our predicted surface brightness profiles for the 
visible (VIS) continuum, with colors and labels as in the left panel. In this case, the shaded 
areas display the regions $450\geq {\rm EW_{H\alpha}\,[\AA] \geq 150}$ and $700\geq {\rm 
EW_{H\alpha}\,[\AA] \geq 50}$. JWST 
observations of the continuum radiation, in the visible wavelength range around  $\sim 6800\,\angs$ 
rest-frame, will enable proving star formation at large distances in the halos observing SFGs. 
Additionally, the VIS profiles will complement the 
UV profiles at large distances, allowing a better comparison of the different continua and line 
profiles which, in turn, unveils the contribution of the different processes yielding LAHs (see 
\S~\ref{sec:sbprofiles}).

\subsubsection{\rm NIRCam {\rm VIS} observations of LAEs}\label{sec:obscont}

We follow the previous observational strategies, using the SNR calculations presented in 
Appendix \ref{sec:snr}, and the instrumental parameters listed in Table \ref{ta:snr}.

   We consider the same sample of LAEs as in \S~\ref{sec:obs}, and the medium-band 
filter F335M, centered at $3.362\, \mu$m and with a bandpass $0.352\, \mu$m, resulting in 
a resolution $R\sim10$. Since this filter is broader than that used to obtain the sample of 
LAEs, the observational depth will be larger, i.e., the number of galaxies falling into the filter band 
is larger than that of LAEs. This (undesired) additional number of galaxies may require the 
modelling of the sources and the removal of extra flux.    

\subsubsection{\rm NIRCam {\rm VIS} observations of SFGs}\label{sec:obs2cont}

   For the observations of star-forming galaxies, we follow the same procedure and sample of 
galaxies as in \S~\ref{sec:obs2}. We use the same filter as above, F335M, noticing that 
the same modelling of sources just discussed will also be necessary in this case.

\begin{figure*}
\includegraphics[width=0.48\textwidth]{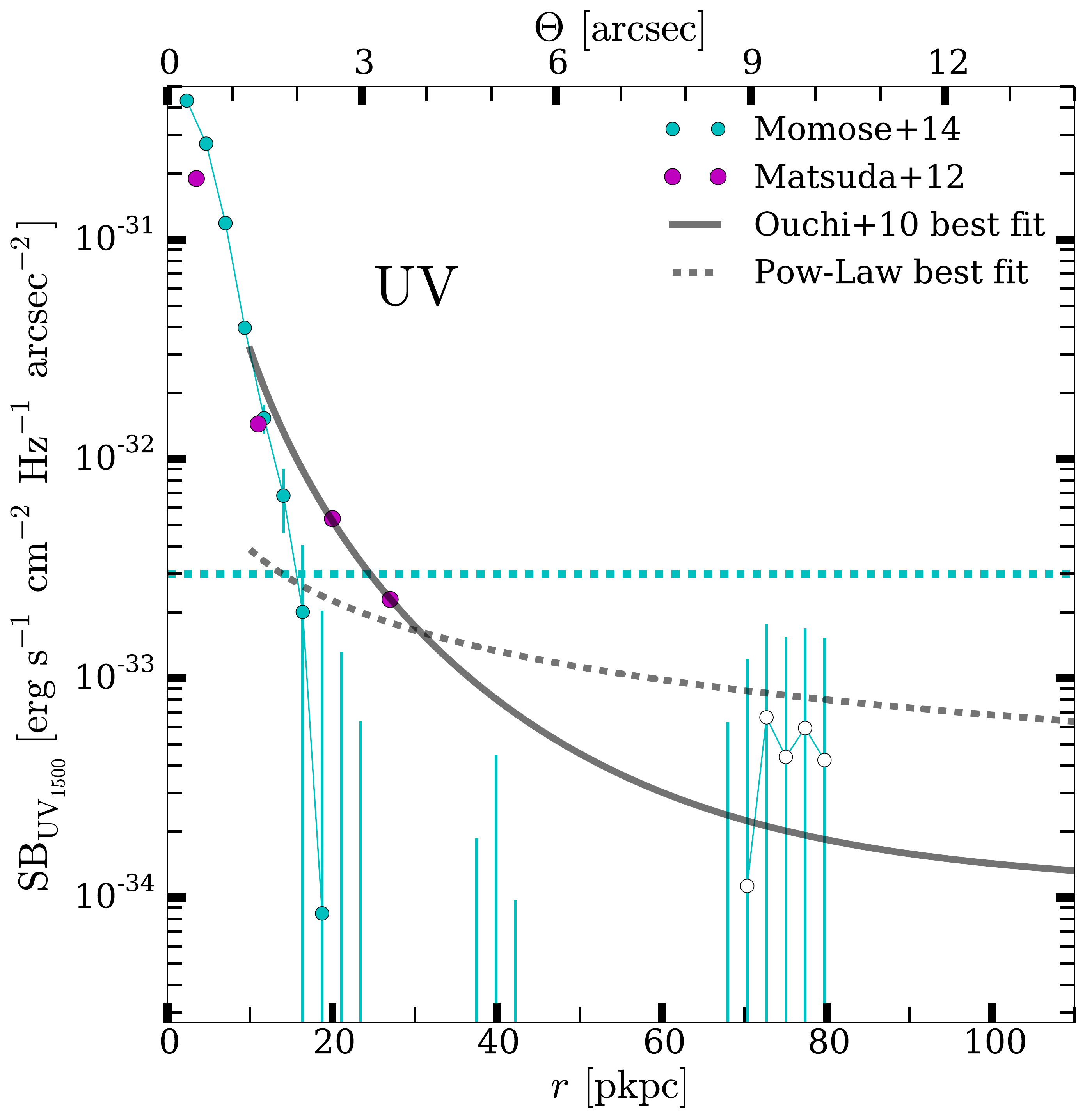}\includegraphics[width=0.48\textwidth]{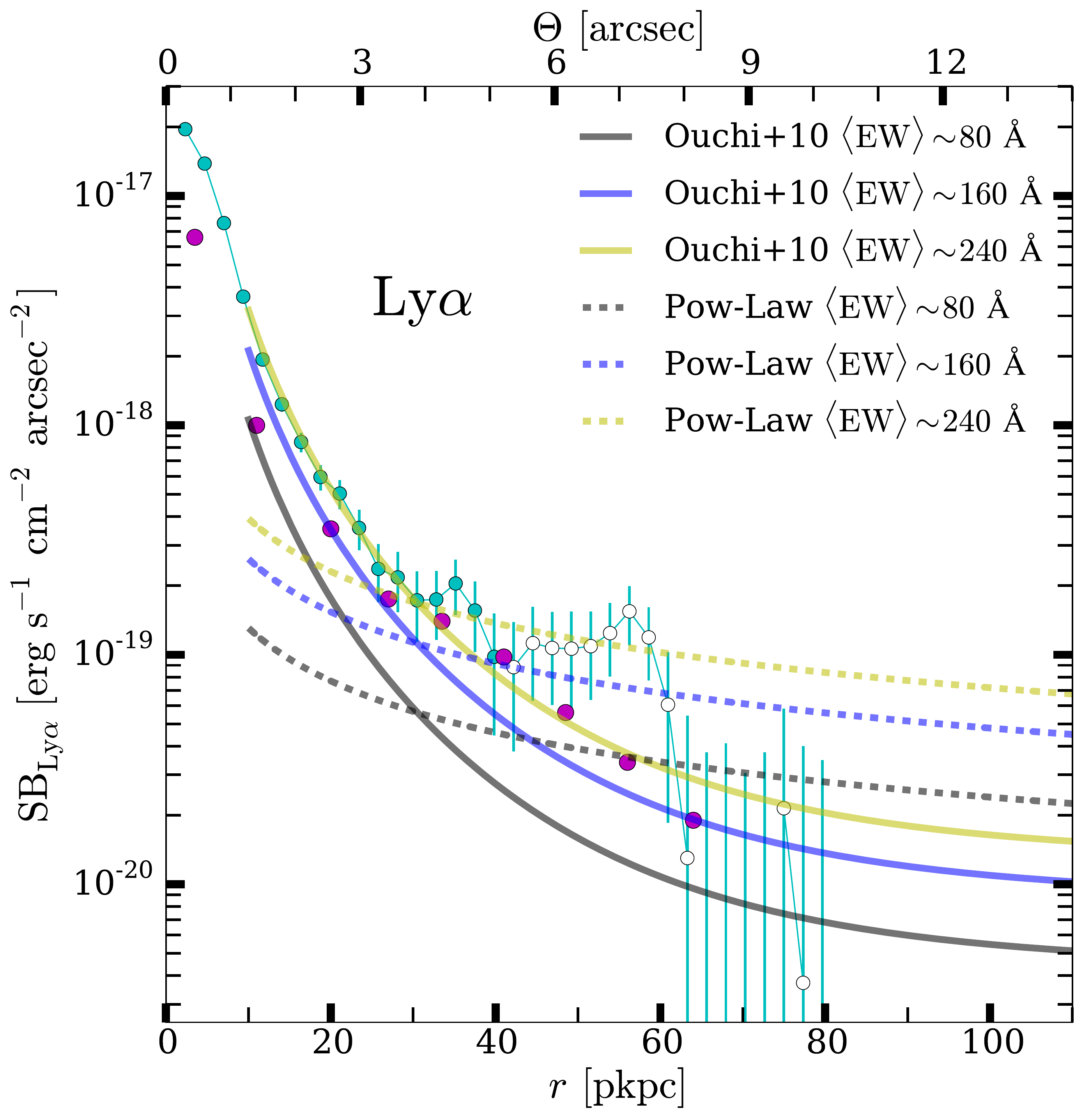}
\caption{{\it Left panel:} Radial UV surface brightness profiles with physical distance from the 
central galaxy for the fiducial ({\it solid grey line}) and the power-law ({\it dashed line}) models, 
reduced by a factor of 3 and 2, respectively, to better fit the data.
{\it Magenta} and {\it cyan dots, line and bars} denote the data as in Figure \ref{fig:sb}. 
{\it Right panel:} Predicted \lya surface brightness profiles for the two models in the left panel 
assuming different population-averaged \lya EW (rest-frame). The profiles with $\langle {\rm EW} 
\rangle \sim 80\, {\rm \AA}$ correspond to the `best-fit' profiles in the left panel assuming the 
same escape fraction for \lya and UV.} 
\label{fig:sbew}
\end{figure*}

\section{Discussion}\label{sec:discussion}

   We discuss below the differences between the parameters for faint satellites and  
brighter galaxies, and the dependence of our results on these values (\S~\ref{sec:cx}). 
In \S~\ref{sec:sbprofiles} we show how the comparison between the {\rm H$\alpha$, Ly$\alpha$}  
and continuum profiles breaks the degeneracies between the different mechanisms that give rise to the 
extended halos.

\subsection{The {\rm EW-PDF}{\rm (}$M_{\rm UV}${\rm )}, {\rm Ly$\alpha$} duty cycle, and 
$C_{{\rm Ly}\alpha}$}\label{sec:cx}

   For any fixed choice of satellite clustering, tuning the model to reproduce the observed 
Ly$\alpha$ surface brightness profile will cause it to overshoot the UV surface brightness profile 
(by a factor of up to $\sim 1.5-3$, depending on the clustering model, see Figure \ref{fig:sb}). 
This effect can be easily remedied by requiring 
that $f_{\rm esc}^{{\rm Ly}\alpha}>f_{\rm esc}^{\rm UV}$. However, resonant scattering typically 
enlarges the total path that Ly$\alpha$ photons travel through dusty, multiphase media, which 
increases the probability that these photons are destroyed by dust grains, relative to that of the 
continuum \citep[see, e.g.,][]{Laursen2013,Gronke2014}. We therefore consider that it is not 
reasonable to require that $f_{\rm esc}^{{\rm Ly}\alpha}>f_{\rm esc}^{\rm UV}$ for the {\it entire 
population}. 

   It is more likely that our adopted conversion factors from star formation rate density 
to integrated volume emissivity ($C_x$ in Eq.~\ref{eq:phi}) differ somewhat. Our current choices 
for $C_{{\rm Ly}\alpha}$ and $C_{\rm UV}$ imply that {\it all} star-forming galaxies produce a 
Ly$\alpha$ line with a rest-frame equivalent width of EW$\sim 80$\angs \hspace{1mm} \citep[see, 
e.g.,][]{Dijkstra2010}, but the EW of the Ly$\alpha$ line can be larger by a factor of a few 
for very young stellar populations \citep[e.g.,][]{Schaerer2003}. $C_x$, especially $C_{{\rm 
Ly}\alpha}$, can be increased for lower metallicity, low SFR galaxies and/or for more top-heavy 
IMFs \citep[][see also the review by \citealt{Kennicutt2012} for \ha and UV]{Raiter2010,
Forero-Romero2013,Masribas2016b}. This interpretation is supported by the short duty-cycle of 
Ly$\alpha$ selected LAEs reported by \cite{Ouchi2010}, which illustrates that the larger EW 
objects are dominated by young stellar populations.

   Our results suggest that, in order to simultaneously reproduce the observed Ly$\alpha$ and UV 
surface brightness profiles, we need the {\it population averaged} rest-frame EW to be $\sim 1.5-3$ 
larger, i.e., we need $\langle {\rm EW} \rangle \sim 120-240$ \AA \hspace{1mm} for \lya. `Population 
averaged' here refers to an average over all satellite galaxies with $M_{\rm UV} \gsim -17$. In the {\it 
left panel} of Figure \ref{fig:sbew}, 
we have reduced the UV surface brightness profiles of the fiducial and power-law models by a factor 
of 3 and 2, respectively, to obtain a good fit to the data. In the {\it right panel} of the same figure, we 
present the corresponding \lya profiles, for different values of $\langle {\rm EW} \rangle$. The 
{\it dashed and solid black lines} denote the power-law and fiducial profiles, respectively, when 
considering the same escape fraction for \lya and UV, as in our previous calculations, i.e., 
$\langle {\rm EW} \rangle \sim 80$ \AA. 
In this case, the profiles fall below the observations as expected. Considering $\langle {\rm EW} 
\rangle \sim 160$ \AA, the fiducial model ({\it solid blue line}) reproduces the data by \cite{Matsuda2012} 
well for $r\lsim30$ pkpc, but is slightly lower at larger distances. Accounting for the contribution of the 
central galaxy, the power-law model ({\it dashed blue line}) may match the data at $r\lsim40$ 
pkpc although is above the observations by \cite{Matsuda2012} at larger distances. If we consider 
$\langle {\rm EW} \rangle \sim 240$ \AA \hspace{1mm} ({\it dashed and solid yellow lines}), 
the models matches the data well at any distance, but if the central galaxy is added, they may  
overpredict the profiles. Therefore, we conclude from this calculation that an average equivalent 
width around $\langle {\rm EW} \rangle \sim 160$ \AA \hspace{1mm} may provide a reasonable 
fit to the data, although the exact value depends on the specific model and contribution of the 
central galaxy.

 \begin{figure} 
\includegraphics[width=0.48\textwidth]{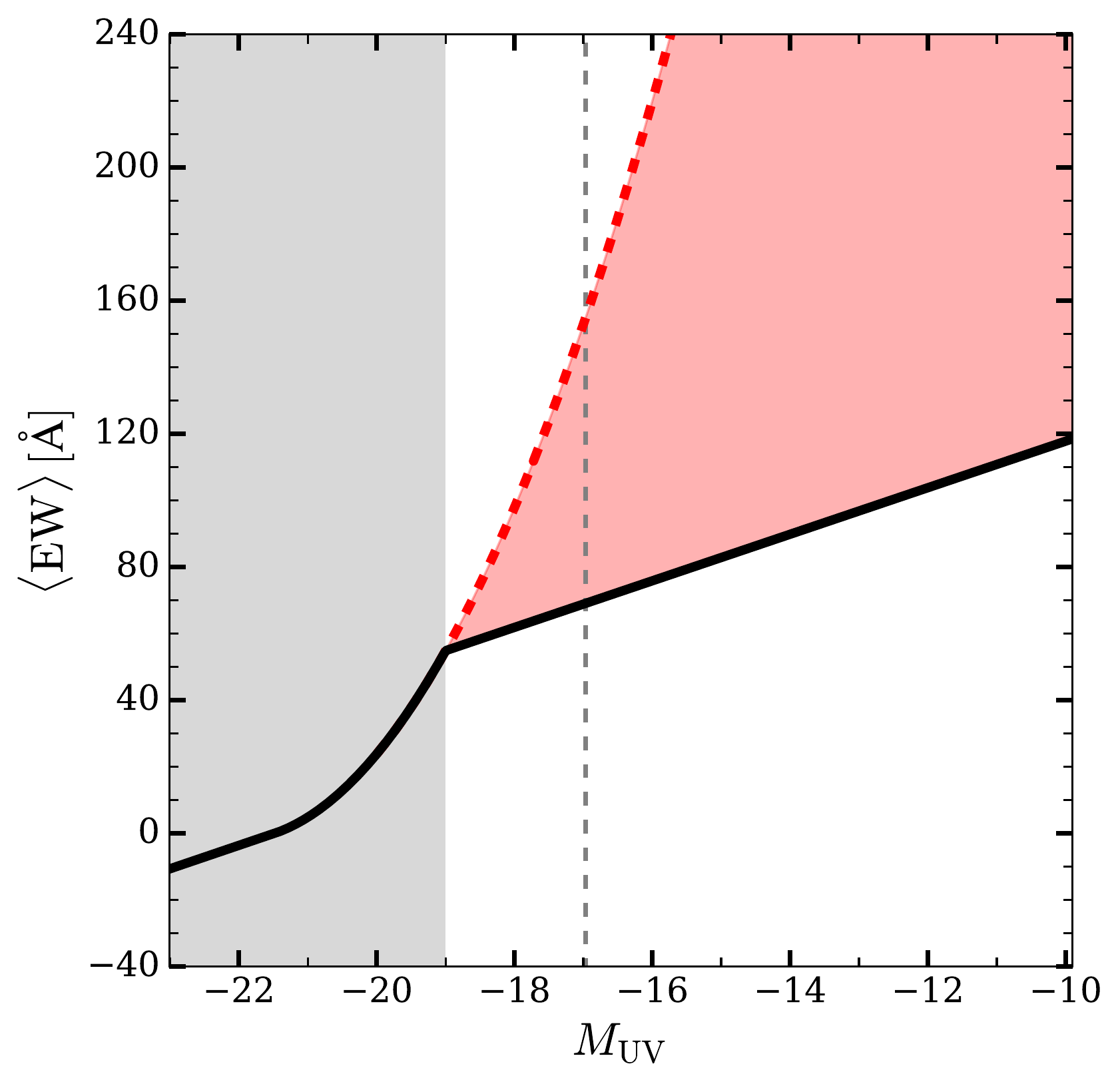}
\caption{Mean \lya rest-frame equivalent width, $\langle{\rm EW} \rangle$, as a function 
of $M_{\rm UV}$, parametrized following the fitting formula by \cite{Dijkstra2012} ({\it solid  
black line}), and extrapolating the evolution observed by \cite{Stark2010} within the range 
$-19\gsim M_{\rm UV} \gsim -22$ ({\it dashed red line}). The {\it red} and the {\it  grey shaded 
areas} represent the regions between the two evolutions and where the model is constrained by 
observations, respectively, and the {\it dashed vertical line} denotes the current observational 
limit, $M_{\rm UV}=-17$, corresponding to $\langle {\rm EW} \rangle = 69$\angs. This figure shows 
that the modest extrapolation of the observed evolution of $\langle{\rm EW} \rangle$ with $M_{\rm UV}$ 
by \cite{Dijkstra2012} can partially account for simultaneously reproducing the observed surface brightness 
profiles of UV and Ly$\alpha$ halos, while steeper evolutions are required for a complete match.} 
\label{fig:EW}
\end{figure}

   Is this $\langle {\rm EW} \rangle$ requirement reasonable? 
There is strong observational support that the Ly$\alpha$ EW-PDF evolves towards fainter 
UV-luminosities. \cite{Dijkstra2012} presented a fitting formula for $P({\rm EW}|M_{\rm UV},z)$ 
constrained by observations.  The {\it solid black line} in Figure~\ref{fig:EW} displays $\langle {\rm 
EW} \rangle$ (rest-frame) as a function of $M_{\rm UV}$ as given by this fitting formula \citep[see ][for 
an alternative parametrization]{Schenker2014}. Figure~\ref{fig:EW} shows that for bright LBGs 
($M_{\rm UV} \lsim -22$, see Figure~A1-A3 of Dijkstra \& Wyithe 2012, which is based in data by 
\citealt{Shapley2003}), $\langle {\rm EW} \rangle \sim 0$. However, $\langle {\rm EW} \rangle$ rapidly 
rises towards lower UV luminosities (based on data by \citealt{Stark2010}) and reaches $\langle {\rm 
EW} \rangle \sim 55$ \AA \hspace{1mm} at $M_{\rm UV} \sim -19$. Due to the lack of observational data, 
Dijkstra \& Wyithe 2012 adopted (conservatively) the same slope as in the range $M_{\rm UV} \le -21.5$ 
for the region $M_{\rm UV} \ge -19$ for the evolution of equivalent width. This modest extrapolation  
can partially account for simultaneously matching the UV and \lya surface brightness profiles. However, 
the evolution may be steeper than assumed by these authors \citep[see, e.g., the recent work at 
$z=7$ by][]{Ota2017}. The {\it dashed red line} in Figure \ref{fig:EW} extrapolates the evolution 
observed by \cite{Stark2010} in the range $-19\gsim M_{\rm UV} \gsim -22$, reaching the 
required value to match both profiles, $\langle {\rm EW} \rangle =240$, quickly after $M_{\rm UV} \sim 
-16$. The {\it shaded red region} shows the area between these two evolutions. 

\subsection{Comparing {\rm H$\alpha$, Ly$\alpha$}, and {\rm Continuum} profiles}\label{sec:sbprofiles}

 Joint analyses of H$\alpha$, Ly$\alpha$ and continuum surface brightness profiles 
are very useful because they enable disentangling the possible origins of the 
extended emission. The differences will rise from the physical 
mechanisms that can yield photons of these three wavelength bands. ({\it i}) Continuum radiation 
is a direct tracer of star formation because it is only produced in the ISM, and it is not 
a resonant transition. ({\it ii}) H$\alpha$ is also produced in the ISM via recombinations 
following hydrogen ionization but, in addition, can be produced far from the star-forming 
regions if ionizing photons reach those distances and ionize the more distant gas (fluorescence).
({\it iii}) Ly$\alpha$ can be produced via the two previous mechanisms, but also by collisional 
excitation of neutral hydrogen accreted into the central galaxy (gravitational cooling). Additionally, 
Ly$\alpha$ is a resonant transition, which allows the Ly$\alpha$ photons to scatter away from the 
sites where they are produced.  The flow chart and plots of Figure \ref{fig:profiles} represent   
a simple method to identify the mechanisms playing a role in the extended emission. The idealized 
diffuse halo in the left part of the figure shows extended \lya ({\it in blue}) but compact \ha and 
continuum emission ({\it in red and green}, respectively). This scenario is a clear indication of  
scattering and/or cooling, as we describe below. The middle plot illustrates 
a more extended \ha halo compared to that of the continuum, indicating that fluorescence is 
important. When star formation occurs far from the center, the continuum will also appear more 
extended, as schematically illustrated in the right plot. \ha and \lya halos will 
also be extended in this case accounting for the nebular radiation of the satellite sources, and can be 
subject to the extra  
contribution of fluorescence, scattering and/or cooling. Additional information can be obtained 
from the radial profiles as follows:

\begin{enumerate}[leftmargin=*]

\item A strong suppression of the continuum and H$\alpha$ surface brightness compared to our 
predictions at a fixed Ly$\alpha$ surface brightness favors the scattering and cooling models. 
Models that purely invoke scattering to explain spatially extended Ly$\alpha$ halos cannot 
produce extended continuum and H$\alpha$ halos. Cooling gives rise to H$\alpha$ and UV  
halos that are suppressed by a factor of $\sim 10$ compared to our predictions here (see 
\citealt{Dijkstra2014} for a review discussing the H$\alpha$ and UV continuum signatures 
of cooling radiation). This scenario corresponds to the 2D plot on the left part of Figure 
\ref{fig:profiles}, where \lya emission appears more extended than the \ha and continuum.

\item Comparing H$\alpha$ and Ly$\alpha$ surface brightness profiles constrains to what extent 
scattering affects the Ly$\alpha$ surface brightness profile. This is because the volume emissivity 
of Ly$\alpha$ and H$\alpha$ closely track each other, while only Ly$\alpha$ photons undergo 
resonant scattering. Scattering systematically flattens the Ly$\alpha$ surface brightness profile, 
as the Ly$\alpha$ photons diffuse outwards prior to escape. 

 \begin{figure*} 
\includegraphics[width=1\textwidth]{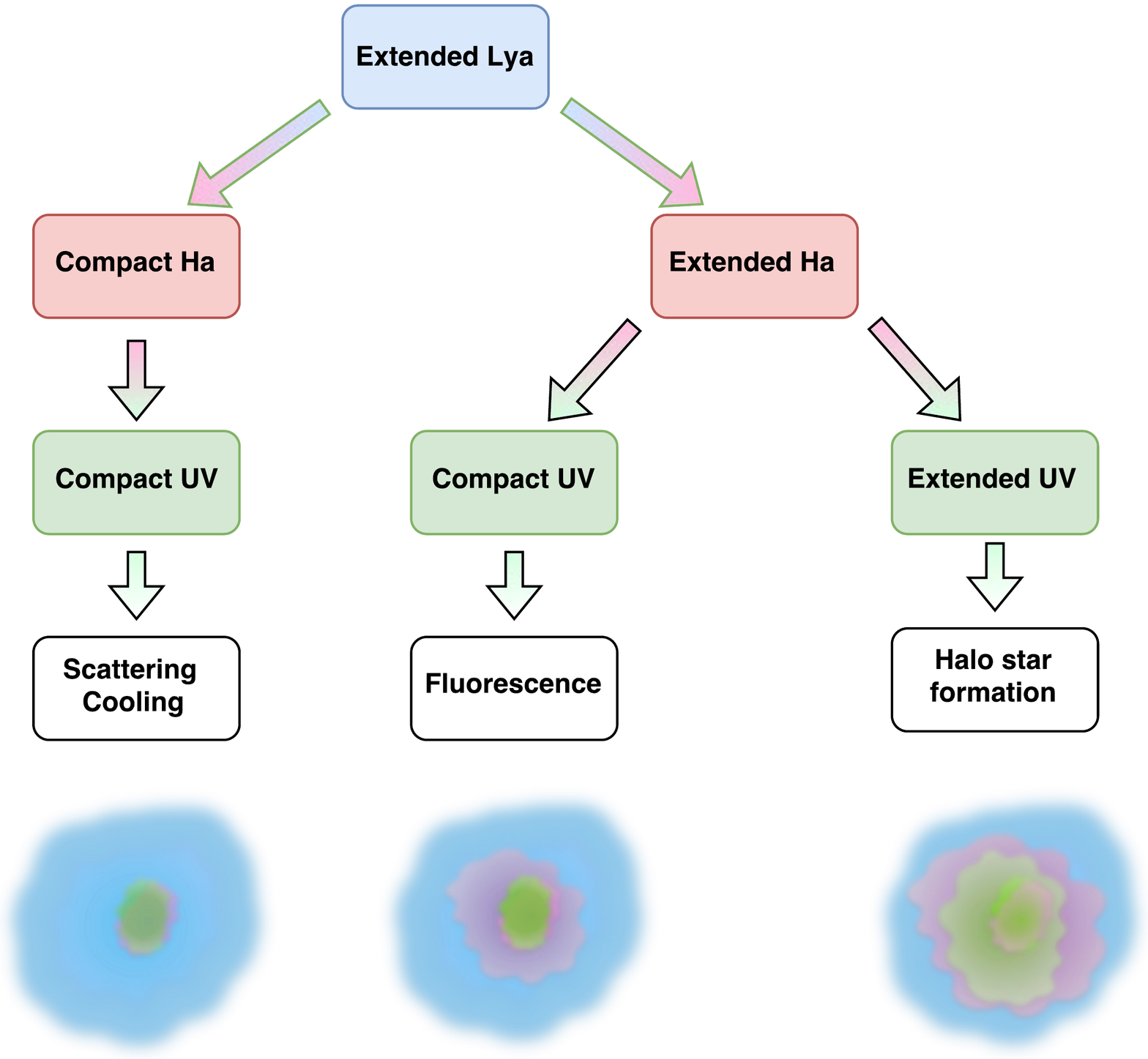}
\caption{Flow chart and plots representing the effect of the different radiative processes on the 
diffuse extended halos. The left plot displays extended \lya emission only {\it in blue}, indicative of 
significant scattering and/or cooling effects. The middle plot shows a larger extent for the \ha emission 
than that of the continuum ({\it red and green, respectively}), implying a contribution of fluorescence. 
The right plot shows extended emission for all cases, indicating the presence of star formation beyond 
the central galaxy.} 
\label{fig:profiles}
\end{figure*}

\item  Comparing H$\alpha$ and continuum surface brightness profiles can determine the importance 
of nebular against fluorescent emission: if no ionizing photons escape from either the central LAE or 
the satellites, then the H$\alpha$ is produced in dense nebulae surrouding O and B stars, and we 
expect the continuum and H$\alpha$ volume emissivity to closely track each other (co-spatial 
{\it green} and {\it red regions} in the left plot of Figure \ref{fig:profiles}). Instead, if ionizing 
photons escape efficiently from low luminosity galaxies \citep[as expected theoretically and 
observationally; e.g.,][see also \citealt{Dijkstra2016} and references therein]{Wise2014,
Japelj2016}, then we expect gas in the CGM to fluoresce in response to the enhanced local ionizing 
radiation field (see \citealt{Masribas2016}). In this case, each satellite galaxy is more extended in 
H$\alpha$ than in the continuum. The resulting overall H$\alpha$ surface brightness profile should also 
be more extended (middle plot in Figure \ref{fig:profiles} where the {\it red region} presents a larger area 
than the {\it green} one).

\end{enumerate}

\section{Conclusions}\label{sec:conclusion}

    We have quantified the contribution of faint (satellite) galaxies ($ M_{\rm UV} > -17$) 
to spatially extended Ly$\alpha$ and UV halos around star forming galaxies. We have applied 
the analytic formalism developed in \cite{Masribas2016} to model the halos around Lyman Alpha 
Emitters (LAEs) at $z=3.1$, for several different satellite clustering prescriptions. 
The predicted surface brightness depends linearly on the product $\bar{\epsilon}^{\rm sat}_x\, 
f_{\rm esc}^x\, b_x$, where $x$ refers to UV, H$\alpha$ and Ly$\alpha$. Here, 
$\bar{\epsilon}^{\rm sat}_x$ denotes the integrated emissivity of faint galaxies, which is directly 
linked to the star formation rate density in these galaxies (see \S~\ref{sec:emiss}), $b_x$ denotes 
the emission bias (see \S~\ref{sec:bias}), and $f^x_{\rm esc}$ denotes the escape fraction (see 
\S~\ref{sec:fesc}). Our main results are as follows:

\begin{itemize}[leftmargin=*]

\item All our models give rise to spatially extended Ly$\alpha$ and UV halos at a level that is 
broadly consistent with observations at $20\lsim r \lsim 40$ pkpc from the centers of LAEs, for 
a reasonable choice of the product $\bar{\epsilon}^{\rm sat}_x\, 
f_{\rm esc}^x\, b_x$. The flatness of the surface brightness profiles depends on the clustering 
prescription at small scales ($r \lsim 100$ pkpc). This result supports the notion that faint 
satellite sources can explain the extended emission, and constrains their presence in the halo 
of more massive galaxies. 

\item For any fixed choice of satellite clustering, the ratio between predicted and observed 
surface brightness at $20\lsim r \lsim 40$ pkpc is higher for UV than for Ly$\alpha$. In other 
words, any given model which we tune to perfectly reproduce the observed Ly$\alpha$ surface 
brightness profile will overshoot the predicted UV surface brightness profile (by a factor of up to 
$\sim 3$, depending on the clustering model). We discussed in \S~\ref{sec:cx} that
this implies that we need the average Ly$\alpha$ EW (rest-frame) of satellite galaxies to lie around  
$\langle {\rm EW} \rangle \sim 120-240$ \AA, the lower end of which is consistent with the 
observed evolution of the Ly$\alpha$ EW-PDF as a function of $M_{\rm UV}$ (see 
Figure~\ref{fig:EW} and Dijkstra \& Wyithe 2012). We found that extrapolating the observed 
evolution of EW with $M_{\rm UV}$ can, at least partially, accommodate these differences. 

\item Because there exist multiple alternative explanations for the presence of extended Ly$\alpha$ 
halos around star forming galaxies (incl.$\,$scattering, cooling, fluorescence; see \S~\ref{sec:intro}), 
it is important to investigate whether there are observables that distinguish between different 
mechanisms. We have therefore also predicted H$\alpha$ surface brightness profiles. Our 
calculations demonstrate that JWST will be able to probe H$\alpha$ surface brightness profiles 
out to distances $r\gsim 80$ pkpc and at levels down to ${\rm SB_{H\alpha}\sim 10^{-21}\,erg\,
s^{-1}\,cm^{-2}\,arcsec^{-2}}$. These \ha observations will enable breaking the 
degeneracies between the different mechanisms that give rise to extended halos.

\end{itemize}

     We generally expect a progressive steepening of the surface 
brightness profiles from \lya to H$\alpha$ to continuum. The exact quantitative steepening depends 
on how efficiently ionizing photons escape from the central galaxy, and its surrounding satellites, 
and also the distribution of self-shielding gas in the CGM of the central galaxy and in the central 
parts of the satellite sources. These more detailed calculations are beyond the scope of our current 
work. Observations of extended halos complement other recently proposed ways to 
constraint escape fractions, such as using H$\beta$ EWs \citep{Zackrisson2013,Zackrisson2016}, 
Ly$\alpha$ line profies \citep{Verhamme2015,Dijkstra2016,Verhamme2016}, 
and covering factor values \citep{Jones2012,Jones2013,Leethochawalit2016,Reddy2016}. 
We will apply our method to investigate the average ionizing escape fraction 
of galaxies during the epoch of cosmic reionization in an upcoming work. 
We also plan to further constrain our modeling by including predictions for the spatial distribution 
(radial offset) of long-duration GRBs from the center of the dark-matter host halo, which will depend 
on the star formation rate and metallicity of the faint satellites \citep{Trenti2015}. 
Current success rates for the detection of GRB host galaxies at $z\sim 3 - 5$ are $\sim 60\%$ 
\citep{Greiner2015}, thus it might be possible that a fraction of the `host-less' GRBs inhabits 
and probes faint (undetected) satellite sources.  

   Our results have focussed on using a LAE as the central galaxy. The reason for using LAEs 
is that there exists good observational data for Ly$\alpha$ halos. However, as Ly$\alpha$ halos 
appear ubiquitously around star forming galaxies \citep{Steidel2011,Wisotzki2015}, our analysis 
can be applied to different populations, allowing for a better understanding 
on the physical processes governing galaxy formation and evolution, and the role played by 
faint, undetected sources to the cosmic photon budget at different epochs.

\section*{acknowledgements}
   LMR is grateful to the ENIGMA group at the MPIA in Heidelberg for kind hospitality and 
inspiring discussions. We thank the anonymous referee for a prompt report and careful 
reading that improved the clarity of our work. Thanks to Joop Schaye and Zolt\'an Haiman 
for comments on the likely distance dependence of the escape fraction, and Nobunari Kashikawa 
for pointing clustering aspects. We thank Joel Primack, David Sobral, kyoung-Soo Lee, 
Lutz Wisotzki, Hakon Dahle, Ainar Drews, Marcia and George Rieke, 
Benjamin Racine, Dan Stark, Johan Fynbo, Eros Vanzella, Jorryt Matthee and Michal 
Michalowski for their comments and suggestions. MD is grateful to the Astronomy 
Department in UCSB for kind hospitality.


\appendix

\section{Appendix A: Distribution of satellite sources}\label{sec:distrib}

   We calculate the number of sources in concentric shells at a distance 
$r$ around the central galaxy for our fiducial clustering model using the expression 
\begin{equation}\label{eq:sch}
n(r) = (1 + z)^3 \int_{r_{\rm min}}^{r_{\rm max}} 4 \pi r^2\, [1+\xi(r)] \,{\rm d}r \int_{L_{\rm min}}^{L_{\rm max}} \phi(L) \, {\rm d}L~,
\end{equation}
where $r_{\rm min}$, $r_{\rm max}$ and $L_{\rm min}$, $L_{\rm max}$ denote the radial 
limits of the shell and luminosity limits, respectively, and 
$\phi(L){\rm d}L$ is the Schechter luminosity function \citep{Schechter1976} 
\begin{equation}
\phi(L){\rm d}L = \phi^* (L/L^*)^{\alpha}\, \exp{L/L^*}\,{\rm d}(L/L^*) ~.
\end{equation}
We run the Poisson distribution on the obtained $n(r)$ values to randomly draw a 
distribution of $10\,000$ integer numbers of galaxies at every concentric shell. 
The parameters of the luminosity functions are quoted in Table \ref{ta:sch}, and are taken 
from the fitting formula by \cite{Kuhlen2012} for UV, and from \cite{Ouchi2008} for 
Ly$\alpha$\footnote{For comparison, we use the two luminosity functions because, 
although the parameters of the UV and \lya luminosity functions are related, this relation 
depends on several assumptions and is not entirely understood \citep[see, e.g., Figure 4 in]
[]{Garel2015}.}. 

\begin{figure*}
\includegraphics[width=0.47\textwidth]{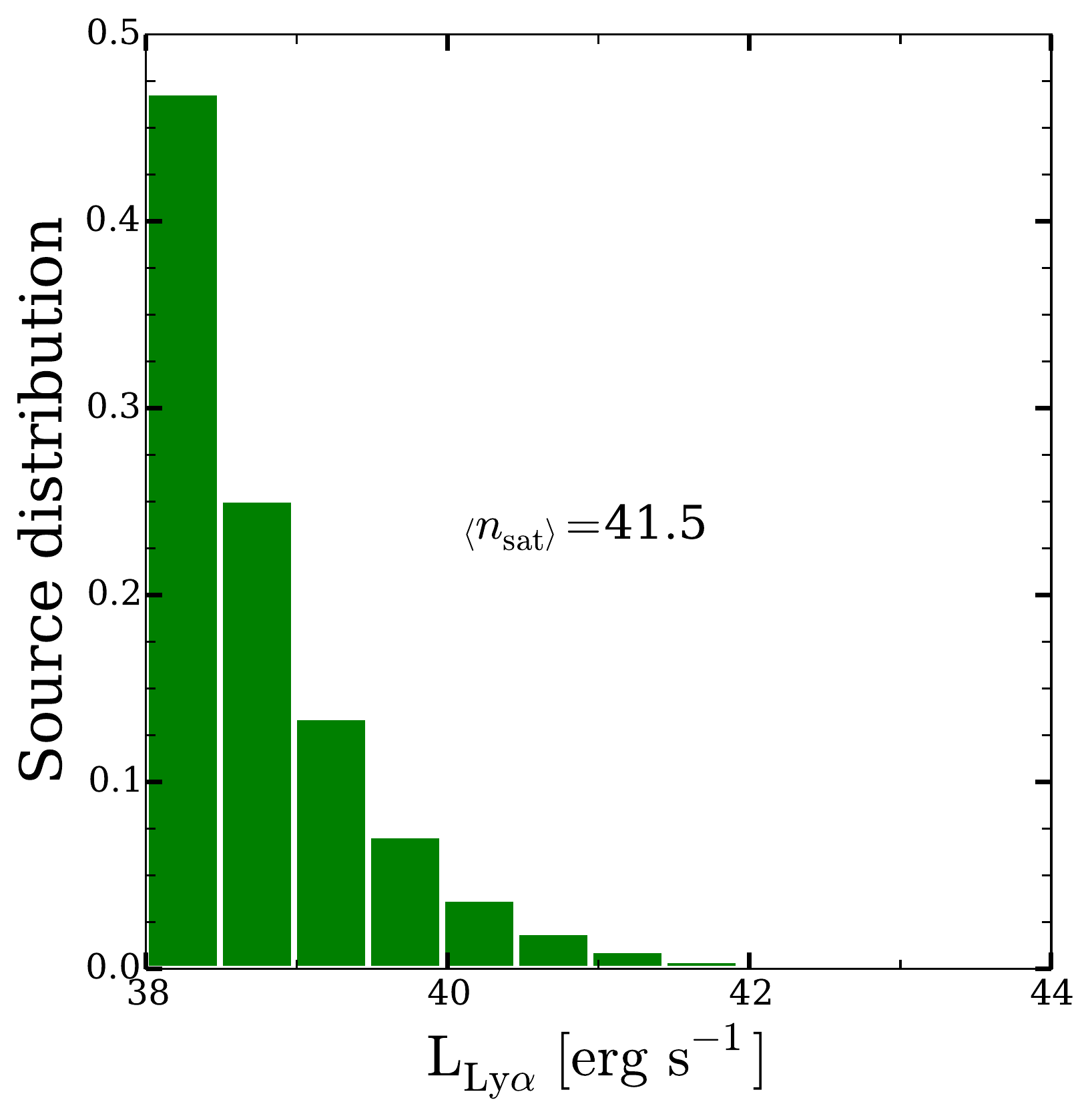}\includegraphics[width=0.42\textwidth]{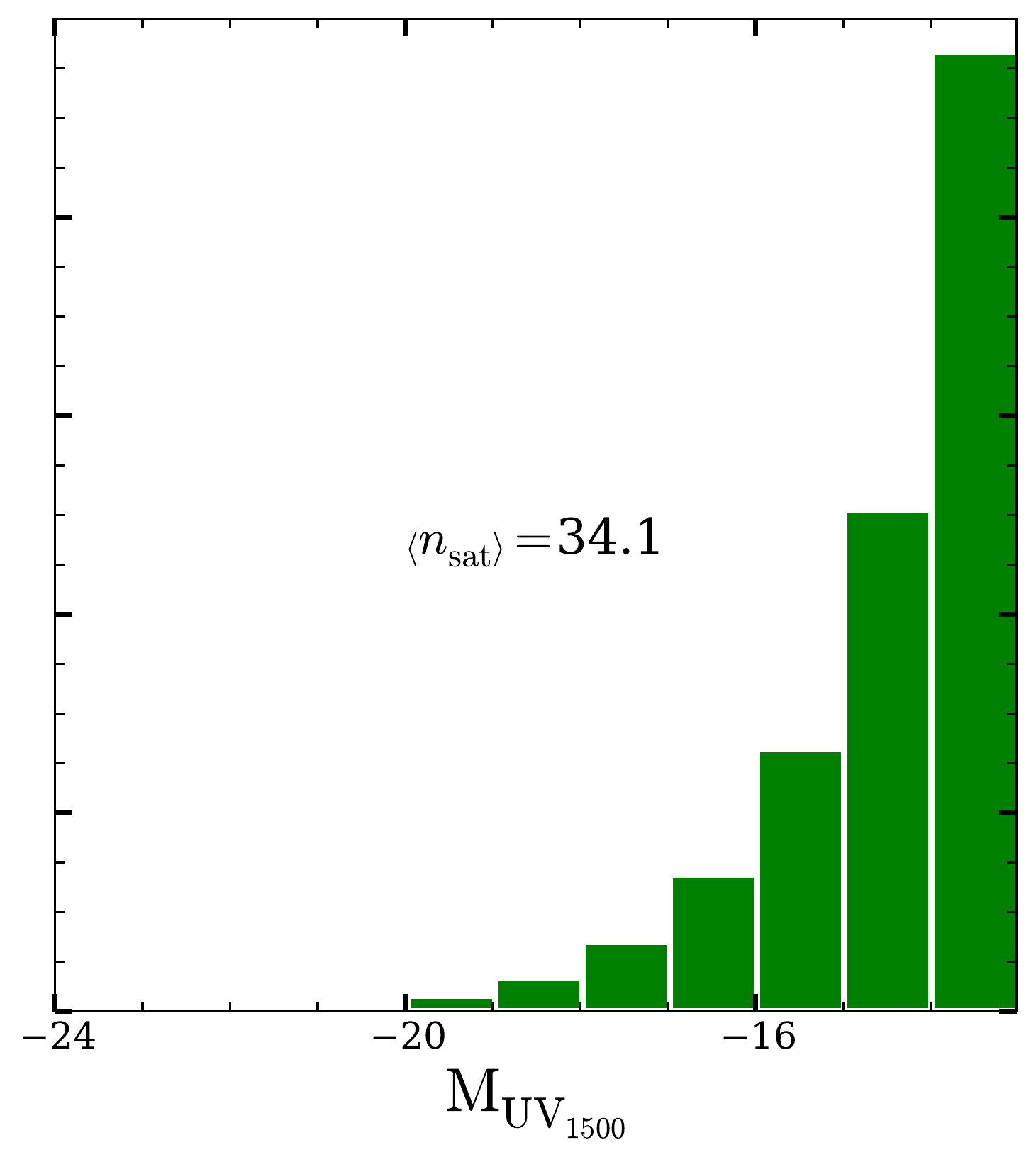}
\caption{Distribution of luminosities for satellite sources in the range $10\leq r \leq 160$ pkpc from the 
central LAE, using the parameters in Table \ref{ta:sch} and our fiducial clustering model. 
{\it Left panel} represents the 
distribution of \lya luminosities and {\it right panel} those of UV magnitudes. Every panel 
quotes the average number of galaxies obtained from Eq.~\ref{eq:sch}. Both panels  
indicate the presence of a few sources close to the observational thresholds, and a 
larger number of significantly fainter objects.} 
\label{fig:distr}
\end{figure*}

   Figure \ref{fig:distr} displays the distribution of sources in the range $10\leq r \leq 160$ 
pkpc, covering the entire region of interest, using the parameters in Table \ref{ta:sch}. 
{\it Left panel} represents the distribution of \lya luminosities and the right one 
of UV magnitudes. Every panel quotes the average number of galaxies from 
Eq.~\ref{eq:sch}, $n(r)$.  Our fiducial model predicts $\sim1-2$ sources in 
the range $L_{Ly\alpha} \sim 10^{40}-10^{41}\,{\rm erg\,s^{-1}}$, the number increasing for fainter 
luminosities. For UV, we obtain $\sim1$ ($\sim2$) sources with magnitudes 
$M_{\rm UV}\sim-17\,(-16)$. We have also computed the distribution of galaxies at different 
radial distances (not shown); the number of galaxies decreases considerably outwards from 
the center, as expected given the profile of the correlation function, and luminosity and 
magnitude distributions display similar profiles than those in Figure \ref{fig:distr}.

   We obtain different emissivity values when integrating the luminosity functions 
compared to those computed previously from the star-formation 
rate density (a factor $\sim 4.6$ reduction for Ly$\alpha$, and $\sim 1.78$ increase for UV). 
These differences may be interpreted as corrections to the adopted values for any of the 
parameters $\bar{\epsilon}_x$, $f_{\rm esc}^x$ and/or $b_x$, but changes in the expressions 
relating emissivity and star formation may (partially or totally) also accommodate such 
differences, since they depend on the adopted IMF and other parameters \citep{Hopkins2006}. 
In addition, we assess below the effect introduced by varying parameters of the 
luminosity functions: 
Setting a minimum \lya luminosity of ${L_{\rm Ly\alpha}}=10^{34}\,  {\rm erg\, s^{-1}}$ and 
an extreme minimum UV magnitude $M_{\rm UV}\sim-3$\, results in emissivity changes by 
less than $1\%$ ($10\%$) for \lya (UV) functions compared to the previous case. 
We obtain, however, thousands of sources in the halo. With lower limits set to   
${L_{\rm Ly\alpha}}=10^{41}\, {\rm erg\, s^{-1}}$ and 
$M_{\rm UV}\sim-17$, the increase of UV emissivity is now lower, a factor $\sim 1.35$, and \lya is 
lower by a factor, $\sim5.45$.  The total average number of sources is $\sim3$ ($\sim1$) for 
UV (Ly$\alpha$), all with luminosities corresponding to the lower limits. 
As mentioned above, several works indicate a steeper \lya faint-end slope than the 
ones in Table \ref{ta:sch} 
\citep{Gronke2015b,Sobral2016}.  Considering $\alpha = - 1.8$ for both functions, the UV 
emissivity is now above by a factor $\sim 2.5$ and \lya below by a factor $\sim 2$ when 
compared to the emissivity from star formation. This result demonstrates that the values 
for the emissivity are more sensible to changes of the faint-end slope than in the 
lower limits of the luminosity functions.

\begin{table}
	\begin{center}
	\caption{\lya and UV luminosity function parameters}	\label{ta:sch}
	\begin{threeparttable}
		\begin{tabular}{cccc} 
		 	 	   			&\lya					&${\rm UV_{1500}}$			&Units \\ \hline
		$ \phi^*$				&$0.92$				&$1.56$					&${\rm (10^{-3}\, Mpc^{-3}\,{\log_{10}{\it L}}^{-1}/UV mag^{-1})}$\\
		$L^*/M^*$\,\tnote{(a)} 	&$5.8\times10^{42}$		&$-20.87$					&${\rm (erg\,s^{-1}/UV\, mag)}$	\\
		$\alpha$				&$-1.50$				&$-1.67$					&\\
		$ L_{\rm Ly\alpha}^{\rm min}/{\rm M_{UV}^{\rm min}}$\,\tnote{(a)}	&$10^{38}$	&$-13$	&${\rm (erg\,s^{-1}/UV\, mag)}$\\
		$ L_{\rm Ly\alpha}^{\rm max}/{\rm M_{UV}^{\rm max}}$\,\tnote{(a)}	&$10^{44}$	&$-24$	&${\rm (erg\,s^{-1}/UV\, mag)}$	\\
		\hline
		\end{tabular}
		\begin{tablenotes}
			\item[(a)] The parameters for \lya are quoted in terms of luminosity and for 
			${\rm UV_{1500}}$ in terms of UV magnitude. 
		\end{tablenotes}		
	\end{threeparttable}
	\end{center}
\end{table}

\section{Appendix B: NIRC\lowercase{am} signal-to-noise calculation}\label{sec:snr}

\begin{table}
	\begin{center}
	\caption{NIRC\lowercase{am} parameters}\label{ta:snr}
	\begin{threeparttable}
		\begin{tabular}{cccc}  
						& {\rm \ha}			&{\rm VIS}		& Units \\\hline
		FOV				&$9.68$				&$9.68$		&${\rm (arcmin^2)}$\\
		${\rm A_{aper}}$ 	&$25$				&$25$		&$({\rm m^2})$	\\
		filter				&F323N				&F335M		&\\
		$\lambda_{\rm obs}\,(z=3.9)$	&$3.237$	&$3.362$ 	&$(\mu$m)\\
		BW				&$0.038$				&$0.352$		&$(\mu$m)\\
		$R$				&$\sim 85$			&$\sim 10$	&\\
		$\eta$			&$0.285$				&$0.458$		&\\
		$t_{\rm exp}$		&$10^{4}$				&$10^4$		&(s)\\ \hline
		\end{tabular}
	\end{threeparttable}
	\end{center}
\end{table}

    We calculate the signal-to-noise ratio in our observations as ${\rm SNR}=N_s / \sqrt{N_s+
N_{\rm sky}}$. We ignore the instrumental noise and systematics since these depend on the 
observational methodology, i.e., number of exposures, number of pixels for source and background 
calculations, rms fluctuations in the detector response after flat-fielding, or the use or not of 
auxiliary calibration data for the dark-current subtraction. Given the large FOV, we expect our noise 
to be dominated by photons instead of systematics. Some of the systematics are 
accounted for in the system throughput parameter $\eta$ and, in any case, we check our results 
with the on-line calculator tool  (see below). $N_s$ and $N_{\rm sky}$ are the azimuthally 
averaged photon counts for the sources and sky, respectively, and are computed as 
\begin{align}
N_s &= \frac{f_{\rm H\alpha}}{h\,\nu_{\rm H\alpha}^{\rm obs}}\, {\rm A_{ aper}}\, \eta\, t_{\rm exp} ~,\\
N_{\rm sky} &= \frac{f_{\rm sky}}{h\,\nu_{\rm H\alpha}^{\rm obs}}\,{\rm BW}\, {\rm A_{ aper}}\, \eta\, t_{\rm exp} ~.
\end{align}

   We have used the on-line JWST Exposure Time Calculator\footnote{\url{https://jwst.etc.stsci.edu/}} 
(ETC) and have found that the results are consistent with our calculations. We find that a line flux 
of $f_{\rm H\alpha}\sim5\times10^{-19}\,{\rm erg\,s^{-1}\,cm^{-2}}$ and $t_{\rm exp}\sim10^4$ s 
correspond to 1$\sigma$ (${\rm SNR=1}$), although this flux can vary by a factor of a few when 
accounting for different readout modes (see, e.g., \url{http://www.stsci.edu/jwst/instruments/nircam/docarchive/JWST-STScI-001721.pdf}).

\bibliographystyle{apj}
\bibliography{lya_emis}\label{References}

\end{document}